\pdfoutput=1
\documentclass{PoS}

\usepackage{amsfonts,amsmath,graphicx,bm,slashed,color}

\newcommand{\bss}[1]{\ensuremath{{\boldsymbol{#1}}}}
\newcommand{\nb}{\phantom{0}}
\newcommand{\wm}{\phantom{-}}

\definecolor{orange}{rgb}{1, 0.55, 0}
\definecolor{brown}{rgb}{0.6, 0.2, 0}
\definecolor{darkblue}{rgb}{0.15,0.15,0.53}
\definecolor{darkgreen}{rgb}{0.0,0.5,0.0}

\title{Flavor physics with $\Lambda_b$ baryons}

\ShortTitle{Flavor physics with $\Lambda_b$ baryons}

\author{\speaker{Stefan Meinel}\\
Center for Theoretical Physics,\\
Massachusetts Institute of Technology,\\
Cambridge, MA 02139, USA\\
E-mail: \email{smeinel@mit.edu}}

\abstract{At the LHC, bottom baryons are being produced in unprecedented quantities, which opens up a new field for
flavor physics. For example, the decay $\Lambda_b \to p\, \mu^- \bar{\nu}$ can be used to obtain a novel determination
of the CKM matrix element $|V_{ub}|$, and the decay $\Lambda_b \to \Lambda\, \mu^+ \mu^-$ probes the weak interactions
at the loop level. The first lattice calculations of the relevant $\Lambda_b \to p$ and $\Lambda_b \to \Lambda$
form factors have recently been performed using domain-wall light quarks and static $b$ quarks. To further reduce
the theoretical uncertainty, one has to go beyond the static approximation. Here I present new calculations of
$\Lambda_b \to p$, $\Lambda_b \to \Lambda$, and $\Lambda_b \to \Lambda_c$ form factors using a relativistic
heavy-quark action.}

\FullConference{31st International Symposium on Lattice Field Theory - LATTICE 2013\\
		July 29 - August 3, 2013\\
		Mainz, Germany}

\begin{document}

\section{Introduction}

\begin{figure}
\begin{center}
\vspace{-2ex}
\includegraphics[width=0.6\linewidth]{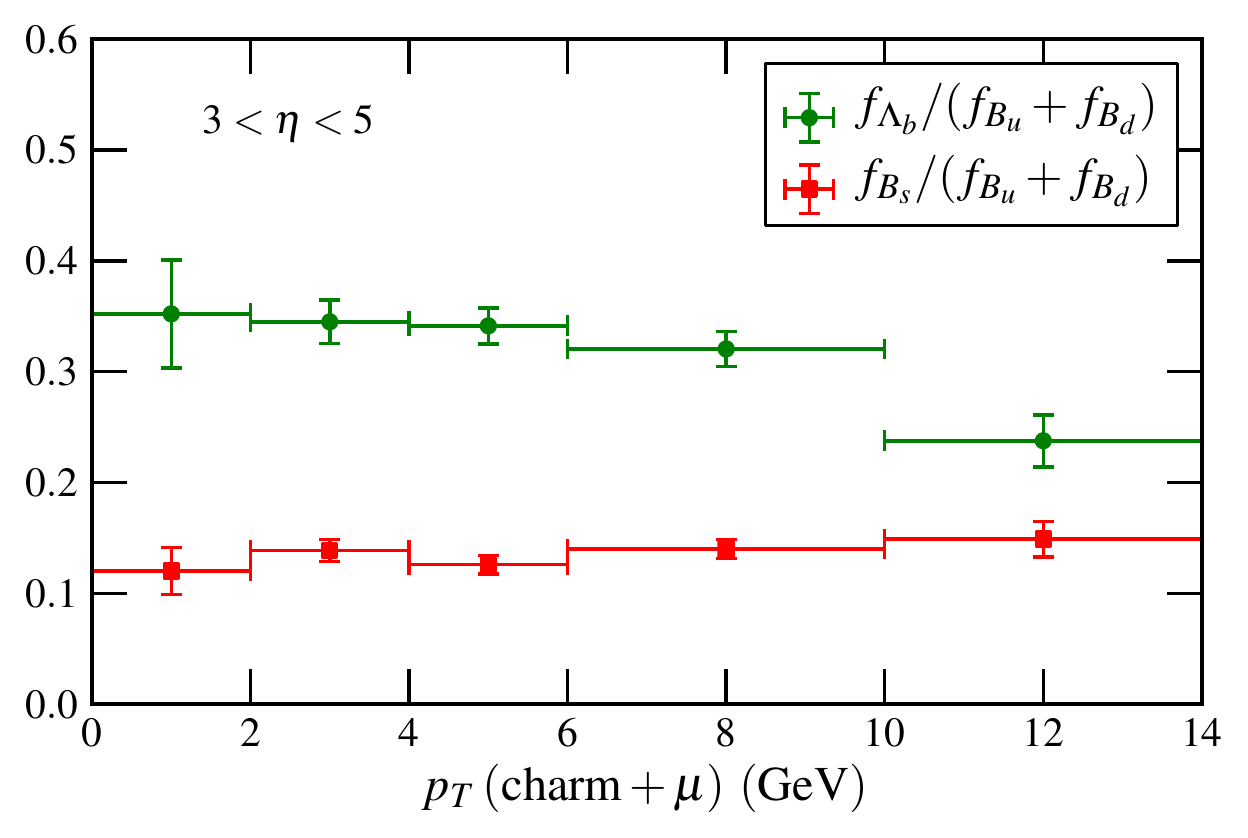}  \vspace{-3ex}
\end{center}
\caption{\label{fig:Lambdabproduction}Production fractions of different species of $b$ hadrons at LHCb,
as a function of transverse momentum \cite{Aaij:2011jp}.}
\end{figure}

Studying the weak interactions of $b$ quarks is one of the most powerful ways to search for physics beyond the Standard Model.
Unlike the dedicated $B$ factories, a high-energy hadron collider produces all species of $b$ hadrons, including baryons. Figure
\ref{fig:Lambdabproduction} shows the relative production fractions observed at LHCb \cite{Aaij:2011jp}. Remarkably,
the number of $\Lambda_b$ baryons is comparable to the number of $B_u$ or $B_d$ mesons, and is significantly higher
than the number of $B_s$ mesons. This raises the questions: How useful are $\Lambda_b$ baryons for flavor physics?
What input is needed from lattice QCD?

In this contribution, I discuss two examples of $\Lambda_b$ decays that offer certain advantages over similar decays
of $B$ mesons: the decay $\Lambda_b \to \Lambda \, \ell^+ \ell^-$, which is a flavor-changing neutral current process
with high sensitivity to new physics, and the decay $\Lambda_b \to p \, \ell^- \bar{\nu}_\ell$, which is a promising
mode for a measurement of the CKM matrix element $|V_{ub}|$ at the Large Hadron Collider. In both cases, form factor
calculations using lattice QCD are needed to utilize the experimental
data.

I begin by discussing the phenomenology of $\Lambda_b \to \Lambda \, \ell^+ \ell^-$ and  $\Lambda_b \to p \, \ell^- \bar{\nu}_\ell$
in Secs.~\ref{sec:LbL} and \ref{sec:Lbp}, highlighting the differences to analogous mesonic decay modes.
In Sec.~\ref{sec:static}, I review the first lattice calculations of $\Lambda_b \to \Lambda$ and $\Lambda_b \to p$
form factors, which were performed with static $b$ quarks \cite{Detmold:2012vy,Detmold:2013nia}. This section
also includes the corresponding results for the $\Lambda_b \to \Lambda \, \ell^+ \ell^-$ and
$\Lambda_b \to p \, \ell^- \bar{\nu}_\ell$ decay observables. In Sec.~\ref{sec:relativistic}, I then present ongoing new calculations
of the complete sets of relativistic $\Lambda_b \to \Lambda$, $\Lambda_b \to p$, and also $\Lambda_b \to \Lambda_c$
form factors. These new calculations will allow predictions of the $\Lambda_b$ decay observables with significantly
reduced uncertainties, as discussed in Sec.~\ref{sec:outlook}.

\section{\label{sec:LbL}The decay $\Lambda_b \to \Lambda \, \ell^+ \ell^-$}

In the decay $\Lambda_b \to \Lambda \, \ell^+ \ell^-$, the bottom quark turns into a strange quark, a process
that in the Standard Model is forbidden at tree-level and proceeds through box and penguin diagrams. This
suppression results in a strong sensitivity to new physics.
At low energy, $b \to s\, \ell^+ \ell^-$ transitions can be described by an effective Hamiltonian
\begin{equation}
\mathcal{H}_{\rm eff} = -\frac{4 G_F}{\sqrt{2}}V_{tb}V_{ts}^* \: \sum_{i} \left[ \: C_i O_i + C_i^{\prime} O_i^{\prime} \: \right], \label{eq:Heff}
\end{equation}
which contains the quark-bilinear operators
\begin{eqnarray}
{O_7^{(\prime)}}    &=& e\: m_b /(16\pi^2)\:\: \bar{s}^a \sigma^{\mu\nu} {P_{R(L)}} b^a \:\:\: F_{\mu\nu}, \label{eq:O7}  \\
{O_8^{(\prime)}}    &=& g\: m_b /(16\pi^2)\:\: \bar{s}^a \sigma^{\mu\nu} {P_{R(L)}} b^b \:\:\: G^{\mu\nu}_{a b}, \label{eq:O8}  \\
{O_9^{(\prime)}}    &=& e^2/(16\pi^2)\:\: \bar{s}^a \gamma^\mu {P_{L(R)}} b^a\:\:\: \bar{\ell} \gamma_\mu \ell, \label{eq:O9}  \\
{O_{10}^{(\prime)}} &=& e^2/(16\pi^2)\:\: \bar{s}^a \gamma^\mu {P_{L(R)}} b^a\:\:\: \bar{\ell} \gamma_\mu \gamma_5 \ell, \label{eq:O10}
\end{eqnarray}
as well as four-quark operators $O_{1,...,6}^{(\prime)}$, including, most importantly,
\begin{equation}
{O_1} = \bar{c}^b\gamma^\mu P_L b^a\:\:\:\bar{s}^a\gamma_\mu P_L c^b, \hspace{4ex}  O_2 = \bar{c}^a\gamma^\mu P_L b^a\:\:\:\bar{s}^b\gamma_\mu P_L c^b.
\end{equation}
The matrix elements of $O_7^{(\prime)}$, $O_9^{(\prime)}$, $O_{10}^{(\prime)}$
factorize into local hadronic matrix elements and trivial leptonic/QED pieces.
The hadronic matrix elements can be written in terms of ten independent form factors,
which are scalar functions of $q^2=(p-p')^2$:
\begin{eqnarray}
 \langle \Lambda |\: \bar{s} \,\gamma^\mu\, b           \:| \Lambda_b \rangle               &=& \bar{u}_\Lambda \big[ {f_1^V}\: \gamma^\mu - {f_2^V}\: i\sigma^{\mu\nu}\! q_\nu/m_{\Lambda_b} + {f_3^V}\: q^\mu/m_{\Lambda_b} \big] u_{\Lambda_b}, \label{eq:fV} \\
 \langle \Lambda |\: \bar{s} \,\gamma^\mu\gamma_5\, b   \:| \Lambda_b \rangle               &=& \bar{u}_\Lambda \big[ {f_1^A}\: \gamma^\mu - {f_2^A}\: i\sigma^{\mu\nu}\! q_\nu/m_{\Lambda_b} + {f_3^A}\: q^\mu /m_{\Lambda_b} \big]\gamma_5\: u_{\Lambda_b}, \label{eq:fA} \\
 \langle \Lambda |\: \bar{s} \,i\sigma^{\mu\nu}\! q_\nu\, b \:| \Lambda_b \rangle           &=& \bar{u}_\Lambda \big[  {f_1^{TV}}\: (\gamma^\mu q^2 - q^\mu \slashed{q} )/m_{\Lambda_b} - {f_2^{TV}}\: i\sigma^{\mu\nu}\! q_\nu  \big] u_{\Lambda_b}, \label{eq:fTV} \\
 \langle \Lambda |\: \bar{s} \,i\sigma^{\mu\nu}\! q_\nu\,\gamma_5\, b \:| \Lambda_b \rangle &=& \bar{u}_\Lambda \big[  {f_1^{TA}}\: (\gamma^\mu q^2 - q^\mu \slashed{q} )/m_{\Lambda_b} - {f_2^{TA}}\: i\sigma^{\mu\nu}\! q_\nu  \big]\gamma_5\: u_{\Lambda_b}. \label{eq:fTA}
\end{eqnarray}
In the Standard Model,
the Wilson coefficients of $O_{7,9,10}$ are (at $\mu=m_b$) $C_7\approx -0.33$, $C_9\approx 4.2$, $C_{10}\approx -4.1$ \cite{Altmannshofer:2008dz},
while the Wilson coefficients of the opposite-chirality operators $O_{7,9,10}^\prime$ are negligibly small.
The other operators in $\mathcal{H}_{\rm eff}$ contribute to the decay via nonlocal hadronic matrix elements containing
an additional insertion of the quark electromagnetic current. At large $q^2$, an operator product expansion can be used
to express these contributions in terms of the local matrix elements (\ref{eq:fV}) and (\ref{eq:fTV}) at leading order \cite{Grinstein:2004vb}.

The first observation of the baryonic decay $\Lambda_b \to \Lambda \, \ell^+ \ell^-$ was reported in 2011 by the CDF
collaboration \cite{Aaltonen:2011qs}. Experimental measurements of mesonic $b\to s$ transitions date back much further
\cite{Ammar:1993sh}. Of the mesonic transitions, the decay $B \to K^* \ell^+ \ell^-$ is particularly useful for individually constraining
all of the Wilson coefficients $C_7$, $C_7^\prime$, $C_9$, $C_9^\prime$, $C_{10}$, and $C_{10}^\prime$. This is because
all of the currents $\bar{s} \sigma^{\mu\nu} b$, $\bar{s} \sigma^{\mu\nu}\gamma_5 b$, $\bar{s} \gamma^\mu b$, and
$\bar{s} \gamma^\mu \gamma_5 b$ have nonzero $B \to K^*$ matrix elements, and these matrix elements depend on the spin of the $K^*$.
In contrast, for the decay $B \to K \ell^+ \ell^-$ with a pseudoscalar kaon, the matrix elements of $\bar{s} \sigma^{\mu\nu}\gamma_5 b$ and
$\bar{s} \gamma^\mu \gamma_5 b$ vanish, so that only the combinations $C_7+C_7^\prime$, $C_9+C_9^\prime$, and $C_{10}+C_{10}^\prime$ are probed.

In the case of $B \to K^* \ell^+ \ell^-$, the $K^*$ immediately undergoes a strong decay, $K^* \to K \pi$, so that the
observed process is actually $B \to K \pi \ell^+ \ell^-$. The angular distribution of this four-body decay can be
analyzed to disentangle the contributions from the various operators in $\mathcal{H}_{\rm eff}$ (see, e.g., Ref.~\cite{Altmannshofer:2008dz}).
This analysis is usually performed in the narrow-width approximation, which assumes an on-shell $K^*$ decaying purely in a
$P$-wave. The narrow-width approximation introduces systematic errors, partly because the observed decay $B \to K \pi \ell^+ \ell^-$
also receives $S$-wave contributions from scalar resonances \cite{Swave}. A complete analysis would require a calculation
of the $B \to K \pi$ (rather than $B \to K^*$) matrix elements of the $b\to s$ currents, which is a very challenging
problem for lattice QCD.

\begin{figure}
\begin{center}
\includegraphics[width=0.4\linewidth]{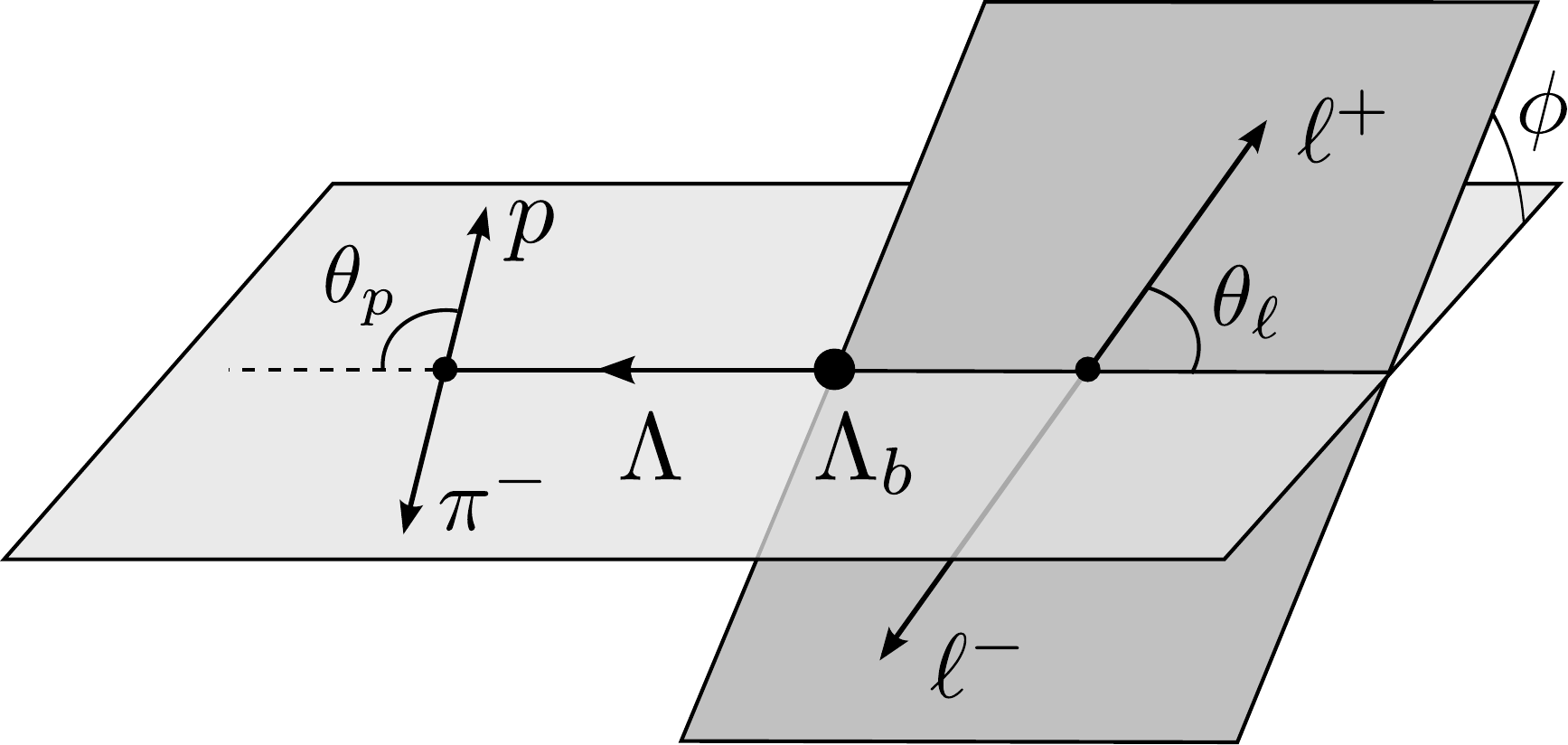}  \vspace{-2ex}
\end{center}
\caption{\label{fig:LambdabLambdaAngular}The decay $\Lambda_b \to \Lambda (\to p \pi^-) \ell^+ \ell^-$. }
\end{figure}

The problem with the narrow-width approximation is completely absent in the baryonic decay $\Lambda_b \to \Lambda \, \ell^+ \ell^-$,
because the $\Lambda$ is stable in QCD. At the same time, the decay $\Lambda_b \to \Lambda \, \ell^+ \ell^-$ is theoretically at least
as powerful as $B \to K^*(\to K \pi) \ell^+ \ell^-$ for probing the full helicity structure of $\mathcal{H}_{\rm eff}$.
The $\Lambda$ in the final state decays through the weak interaction into $n\,\pi^0$ or $p\, \pi^-$ (the latter mode
is reconstructed in experiments). The kinematics of the two-stage decay $\Lambda_b \to \Lambda (\to p\,\pi^-) \ell^+ \ell^-$
can be described by four variables: the angles $\theta_p$, $\theta_l$, $\phi$ (see Fig.~\ref{fig:LambdabLambdaAngular})
and the invariant mass squared of the lepton pair, $q^2$. Expressions for the four-fold decay distribution
(for the case of unpolarized $\Lambda_b$) can be found in Ref.~\cite{Gutsche:2013pp}. For example, after integrating over $\theta_l$ and $\phi$,
the $\theta_p$-dependence of the decay rate is
\begin{equation}
 \frac{\mathrm{d}\Gamma\big[\Lambda_b\to\Lambda(\to p\,\pi^-)\ell^+\ell^-\big]}{\mathrm{d}q^2\:\mathrm{d}\cos{\theta_p}}
 =\mathcal{B}(\Lambda \to p\pi^-)\frac12 \frac{\mathrm{d}\Gamma\big[\Lambda_b\to\Lambda\ell^+\ell^-\big]}{\mathrm{d}q^2}
 \left(1+a\: P_z^{(\Lambda)}\: \cos \theta_p \right), \label{eq:thetap}
\end{equation}
where $P_z^{(\Lambda)}(q^2)$ is the $z$-component of the $\Lambda$ polarization \cite{Gutsche:2013pp, Huang:1998ek} and
$a=0.642(13)$ \cite{Beringer:1900zz} is the parity-violating ``analyzing power'' of $\Lambda \to p\,\pi^-$. Equation (\ref{eq:thetap})
shows that the decay  $\Lambda_b \to \Lambda (\to p\,\pi^-) \ell^+ \ell^-$ features a hadron-side forward-backward asymmetry,
which is not present in $B \to K^*(\to K \pi) \ell^+ \ell^-$. Furthermore, the $\phi$-dependence of the decay distribution
is sensitive to the $CP$-odd transverse polarization $P_x^{(\Lambda)}$ \cite{Gutsche:2013pp, Chen:2002rg}.

A related rare decay is the radiative mode $\Lambda_b \to \Lambda \gamma$, which is mediated primarily by the operators
$O_7$ and $O_7^\prime$. The possibility of using the $\Lambda_b$ and/or $\Lambda$ polarizations to disentangle the
contributions of $O_7$ and $O_7^\prime$ was discussed for example in Refs.~\cite{Mannel:1997xy, Huang:1998ek, Hiller:2001zj}.

\section{\label{sec:Lbp}The decay $\Lambda_b \to p \, \ell^- \bar{\nu}_\ell$}

\begin{figure}
\begin{center}
\vspace{-1ex}
\includegraphics[width=0.77\linewidth]{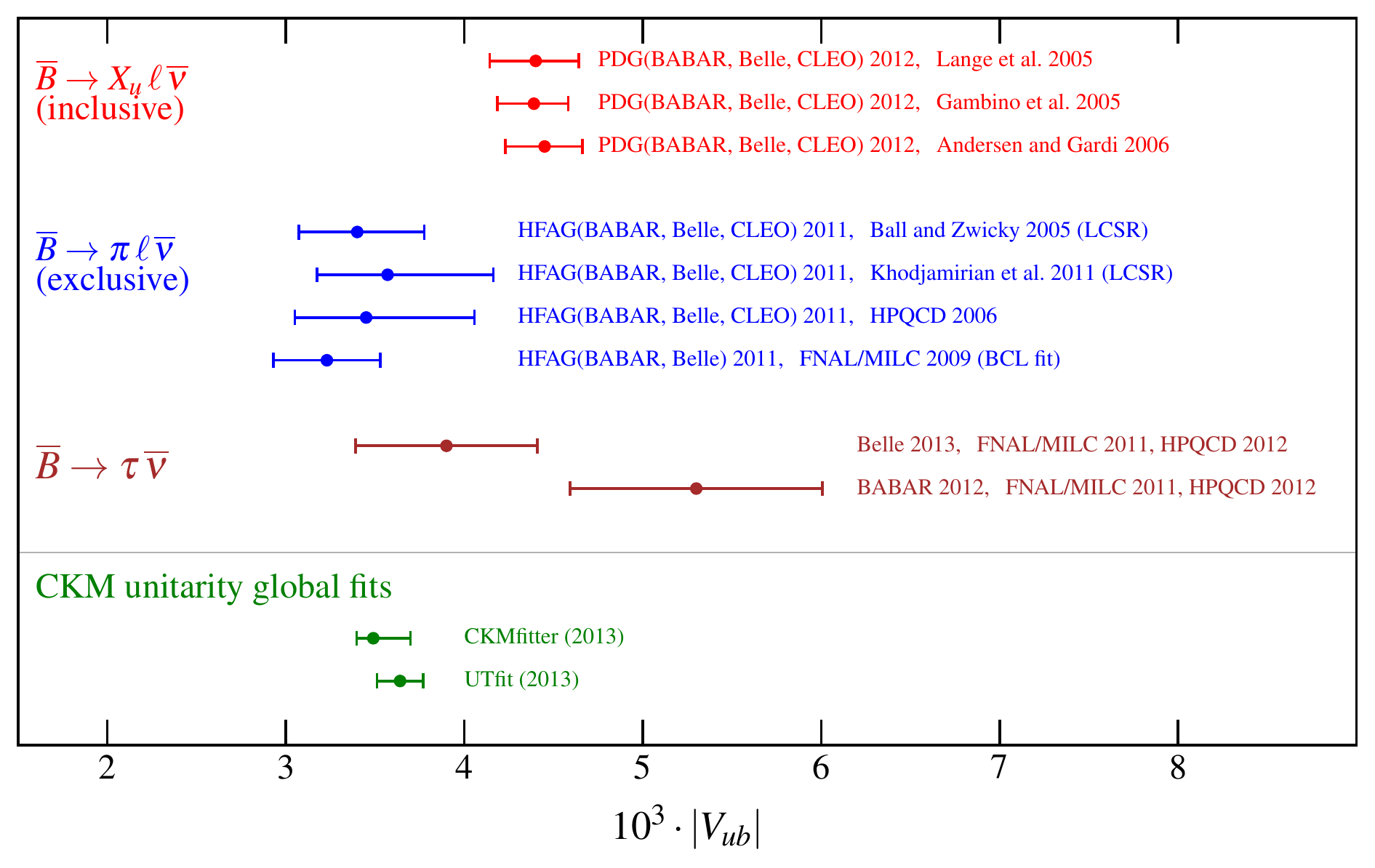}  \vspace{-3ex}
\end{center}
\caption{\label{fig:Vub}Summary of results for $|V_{ub}|$ as of July 2013 \cite{Beringer:1900zz,Amhis:2012bh,FLASY2013,CKMfits}.}
\vspace{1ex}
\end{figure}

The rate of the decay $\Lambda_b \to p \, \ell^- \bar{\nu}_\ell$, where the bottom quark turns into an up quark,
is proportional to $|V_{ub}|^2$; this decay can therefore provide a novel measurement of the poorly known magnitude
of the CKM matrix element $V_{ub}$. So far, all measurements of $|V_{ub}|$ have used $B$ meson decays and were performed
at dedicated $B$ factories. As shown in Fig.~\ref{fig:Vub}, there is a significant discrepancy between the most precise
extraction from exclusive $\bar{B} \to \pi \ell \bar{\nu}$ decays \cite{Amhis:2012bh}, which uses $B\to\pi$ form factors
from lattice QCD \cite{Bailey:2008wp}, and the determinations from inclusive $\bar{B} \to X_u \ell \bar{\nu}$ decays
\cite{Beringer:1900zz}. The results from the purely leptonic decay $\bar{B} \to \tau \bar{\nu}_\tau$ have large
experimental uncertainties due to limited statistics.

Given that the $B$ factories are no longer running, it is highly desirable to perform an independent
measurement of $|V_{ub}|$ at the LHC. One reason why measurements of $\bar{B} \to \pi \ell \bar{\nu}$
are difficult at the LHC is the large pion background produced in a hadron collider. The final state
of $\Lambda_b \to p \, \ell^- \bar{\nu}_\ell$ is more distinctive, making this the preferred mode for
a $|V_{ub}|$ determination with the LHCb experiment \cite{Egede}. The data analysis for $\Lambda_b \to p \, \mu^- \bar{\nu}_\mu$
is currently in progress \cite{Egede}. 

There is also another reason why the decay $\Lambda_b \to p \, \ell^- \bar{\nu}_\ell$ is very interesting.
The discrepancy between the $\bar{B} \to \pi \ell \bar{\nu}$ and $\bar{B} \to X_u \ell \bar{\nu}$ determinations could in principle
be caused by new physics that introduces a right-handed current, with a new coefficient $V_{ub}^\prime$,
in the $b\to u\, \ell^- \bar{\nu}_\ell$  effective Hamiltonian \cite{VubRH}: \vspace{-1ex}
\begin{equation}
\mathcal{H}_{\rm eff}=\frac{G_F}{\sqrt{2}}\Big[V_{ub} (\bar{u}\gamma_\mu b - \bar{u}\gamma_\mu \gamma_5 b)
+ V_{ub}^\prime (\bar{u}\gamma_\mu b + \bar{u}\gamma_\mu \gamma_5 b) \Big] (\bar{\ell}\gamma_\mu \nu - \bar{\ell}\gamma_\mu \gamma_5 \nu). \label{eq:VubHeff}
\end{equation}
A non-zero $V_{ub}^\prime$ could explain the order
$|V_{ub}^{\rm eff}|_{\bar{B} \to \pi \ell \bar{\nu}} < |V_{ub}^{\rm eff}|_{\bar{B} \to X_u \ell \bar{\nu}} < |V_{ub}^{\rm eff}|_{\bar{B} \to \ell \bar{\nu}}$
of extracted values, because these three decays depend on different combinations of the vector and axial vector currents (see Table \ref{tab:Vubcurrents}).
The baryonic decay $\Lambda_b \to p \, \ell^- \bar{\nu}_\ell$ receives contributions from both the vector and axial vector currents, and
will provide a valuable test of the new-physics scenario (\ref{eq:VubHeff}).

\begin{table}
\begin{center}
\begin{tabular}{lcc}
\hline\hline
 Process & $\bar{u}\gamma_\mu b$ & $\bar{u}\gamma_\mu \gamma_5 b$ \\
\hline
 $\bar{B}\phantom{^{2^2}}\hspace{-1.6ex} \to \pi\, \ell\, \bar{\nu}_\ell$          & \textcolor{darkgreen}{$\checkmark$}  & \textcolor{red}{$\times$}            \\
 $\bar{B}\phantom{^{2^2}}\hspace{-1.6ex} \to \ell\, \bar{\nu}_\ell$                & \textcolor{red}{$\times$}            & \textcolor{darkgreen}{$\checkmark$}  \\
 $\bar{B}\phantom{^{2^2}}\hspace{-1.6ex} \to X_u\, \ell\, \bar{\nu}_\ell$          & \textcolor{darkgreen}{$\checkmark$}  & \textcolor{darkgreen}{$\checkmark$}  \\
 $\Lambda_b \to p\, \ell\, \bar{\nu}_\ell$                                         & \textcolor{darkgreen}{$\checkmark$}  & \textcolor{darkgreen}{$\checkmark$}  \\
\hline\hline
\end{tabular} \vspace{-2ex}
\end{center}
\caption{\label{tab:Vubcurrents}Currents contributing to different $b\to u \ell\, \bar{\nu}_\ell$ processes.}
\end{table}

The $\Lambda_b \to p$ form factors are defined as in Eqs.~(\ref{eq:fV}-\ref{eq:fTA}), with the appropriate replacements.
With the effective Hamiltonian (\ref{eq:VubHeff}), the $\Lambda_b \to p \, \ell^- \bar{\nu}_\ell$ decay rate depends on the six
form factors $f_{1,2,3}^V$ and $f_{1,2,3}^A$ (in the approximation $m_\ell=0$, the form factors $f_3^V$ and $f_3^A$ do not contribute).

\section{\label{sec:static}Lattice calculation with static $b$ quarks}

The expressions (\ref{eq:fV})-(\ref{eq:fTA}) for the decomposition of the $\Lambda_b \to \Lambda$ or $\Lambda_b \to p$
matrix elements in terms of form factors simplify dramatically if the $b$ quark is treated at leading order in heavy-quark
effective theory (HQET), i.e. in the static limit. The matrix elements with arbitrary gamma matrices in the current can then be written
in terms of only two form factors $F_1$ and $F_2$ as follows:
\begin{equation}
\langle X |\: \bar{q}\, {\Gamma} Q \: | \Lambda_Q \rangle \:\:=\:\: \bar{u}_X \: \left[{F_1}^{(\Lambda_Q \to X)}
+ \slashed{v}\: {F_2}^{(\Lambda_Q \to X)} \right] \: {\Gamma} \: u_{\Lambda_Q}, \label{eq:staticFF}
\end{equation}
where $X=\Lambda,p$ and $q=s,u$. Here, $v$ is the four-velocity of the $\Lambda_Q$ baryon, and $Q$ is the static heavy-quark
field satisfying $\slashed{v}Q=Q$. The form factors $F_1$ and $F_2$ are functions of $v\cdot p'$, where $p'$ is the momentum
of the final-state baryon (in the $\Lambda_Q$ rest frame, one has $v\cdot p'=E_X$). The static approximation is accurate
up to corrections of order $\Lambda/m_b$, where $\Lambda$ is the typical momentum scale of the light degrees of freedom
(at zero recoil, $\Lambda\sim\Lambda_{\rm QCD}$).

Lattice QCD determinations of the form factors $F_1$ and $F_2$ are published in Ref.~\cite{Detmold:2012vy} for $\Lambda_Q \to \Lambda$ and
in Ref.~\cite{Detmold:2013nia} for $\Lambda_Q \to p$. We performed these calculations with domain-wall $u$, $d$, and $s$
quarks, using gauge field configurations generated by RBC/UKQCD \cite{Aoki:2010dy}. Their parameters are shown in
Table \ref{tab:params}. We implemented the static heavy quark using the Eichten-Hill action \cite{Eichten:1989kb} with
one level of HYP smearing \cite{DellaMorte:2005yc}. In addition to reducing the number of form factors and hence simplifying
the data analysis, the static limit also saves computer time, because the heavy-quark propagators can be constructed
without any inversions. The $\mathcal{O}(a)$-improved matching of the static-light currents from lattice HQET to continuum
HQET was done at one loop in mean-field-improved perturbation theory \cite{Ishikawa:2011dd}.

\begin{table}
\begin{center}
\small
\begin{tabular}{ccccccccccc}
\hline\hline
Set & $N_s^3\times N_t$ & $\beta$ & $am_{s}^{(\mathrm{sea})}$  & $am_{u,d}^{(\mathrm{sea})}$
& $a$ (fm) & $am_{s}^{(\mathrm{val})}$ & $am_{u,d}^{(\mathrm{val})}$  & \hspace{-1.5ex} $m_\pi^{(\mathrm{vv})}\!$ (MeV) \hspace{-1.5ex}  & \hspace{-1ex} $m_{\eta_s}^{(\mathrm{vv})}\!$ (MeV) \hspace{-1.5ex} \\
\hline
$\mathtt{C14}$ & $24^3\times64$ & $2.13$ & $0.04$ & $0.005$ & $0.1119(17)$ & $0.04$ & $0.001$    & 245(4)   & 761(12)  \\
$\mathtt{C24}$ & $24^3\times64$ & $2.13$ & $0.04$ & $0.005$ & $0.1119(17)$ & $0.04$ & $0.002$    & 270(4)   & 761(12)  \\
$\mathtt{C54}$ & $24^3\times64$ & $2.13$ & $0.04$ & $0.005$ & $0.1119(17)$ & $0.04$ & $0.005$    & 336(5)   & 761(12)  \\
$\mathtt{C53}$ & $24^3\times64$ & $2.13$ & $0.04$ & $0.005$ & $0.1119(17)$ & $0.03$ & $0.005$    & 336(5)   & 665(10)  \\
$\mathtt{F23}$ & $32^3\times64$ & $2.25$ & $0.03$ & $0.004$ & $0.0849(12)$ & $0.03$ & $0.002$    & 227(3)   & 747(10)  \\
$\mathtt{F43}$ & $32^3\times64$ & $2.25$ & $0.03$ & $0.004$ & $0.0849(12)$ & $0.03$ & $0.004$    & 295(4)   & 747(10)  \\
$\mathtt{F63}$ & $32^3\times64$ & $2.25$ & $0.03$ & $0.006$ & $0.0848(17)$ & $0.03$ & $0.006$    & 352(7)   & 749(14)  \\
\hline\hline
\end{tabular}\vspace{-2ex}
\end{center}
\caption{\label{tab:params} Parameters of the gauge configurations and light-quark propagators.}
\end{table}

\begin{figure}[!ht]
\begin{center}
\includegraphics[width=0.49\linewidth]{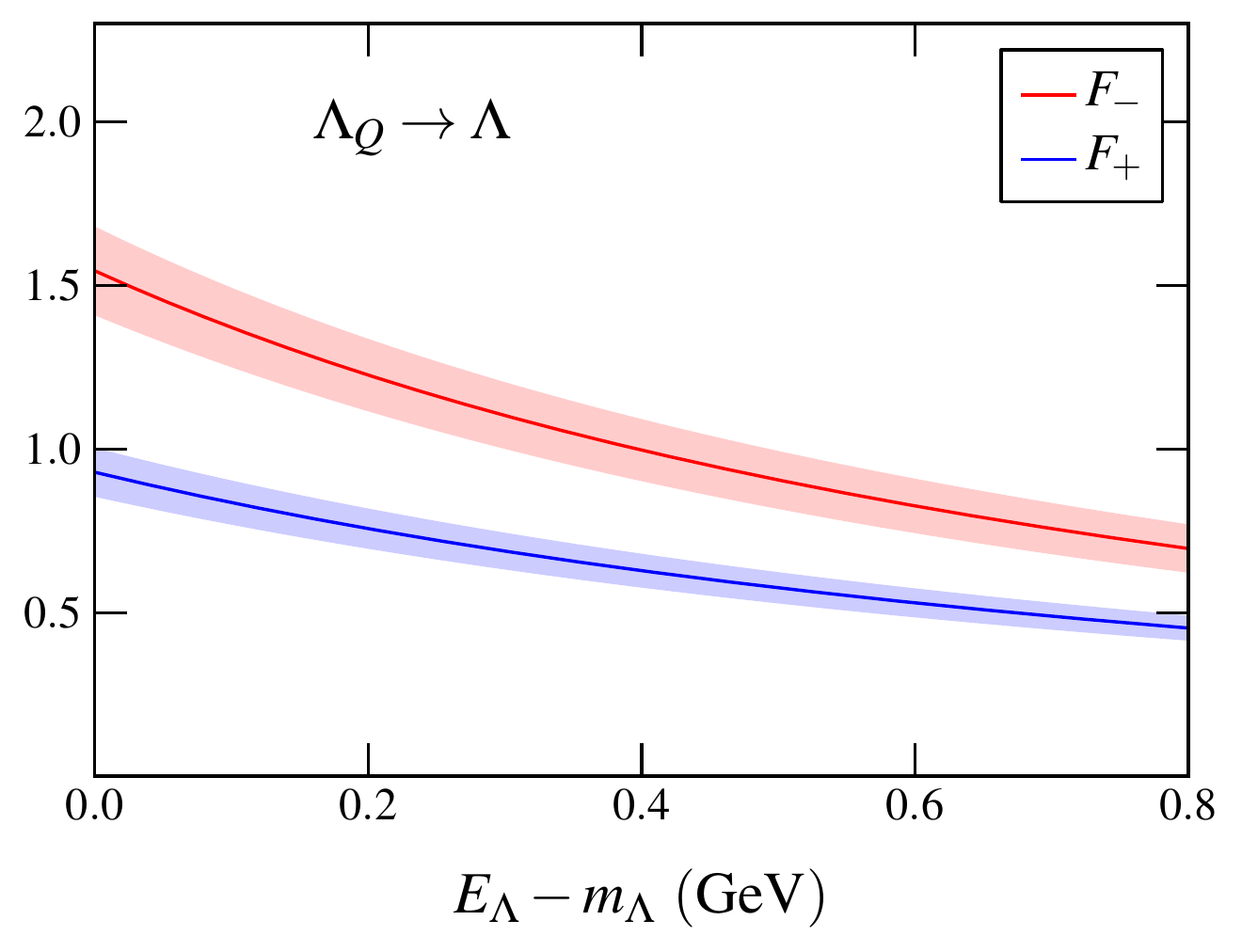}
\hfill \includegraphics[width=0.49\linewidth]{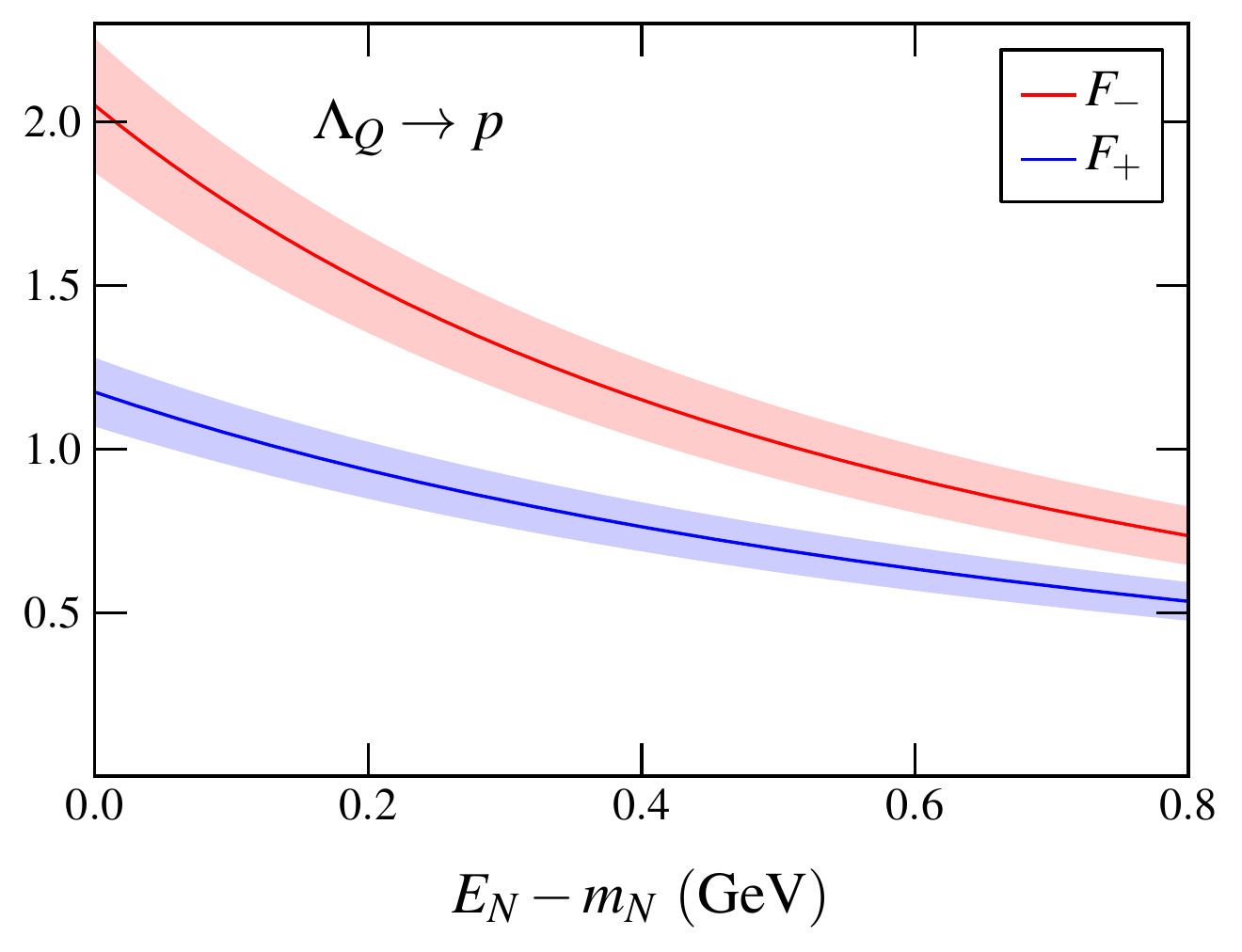}  \vspace{-3ex}
\end{center}
\caption{\label{fig:staticFFs}Final results for the HQET form factors $F_\pm=F_1\pm F_2$ for $\Lambda_Q \to \Lambda$ (left) and $\Lambda_Q \to p$ (right).
The shaded bands show the total (including 8\% systematic) uncertainties.}
\end{figure}

In the data analysis, it turned out to be more natural to work with the linear combinations
$F_+=F_1+F_2$ and $F_-=F_1-F_2$. The final results for these form factors, extrapolated to the continuum
limit and to the physical light-quark masses, are shown in Fig.~\ref{fig:staticFFs}. These extrapolations
assumed a linear dependence on the valence quark masses, a quadratic dependence on the lattice spacing,
and used a dipole model to interpolate the dependence on $E_X-m_X$. Within the statistical uncertainties,
the results do not actually show any significant dependence on the quark masses or lattice spacing, and the fits
have good quality. We estimated the total systematic uncertainties in the form factors $F_+$ and $F_-$ to be
8\% \cite{Detmold:2012vy, Detmold:2013nia}, dominated by the uncertainty associated with the perturbative current
matching.

\begin{figure}[!ht]
\begin{center}
\includegraphics[width=0.55\linewidth]{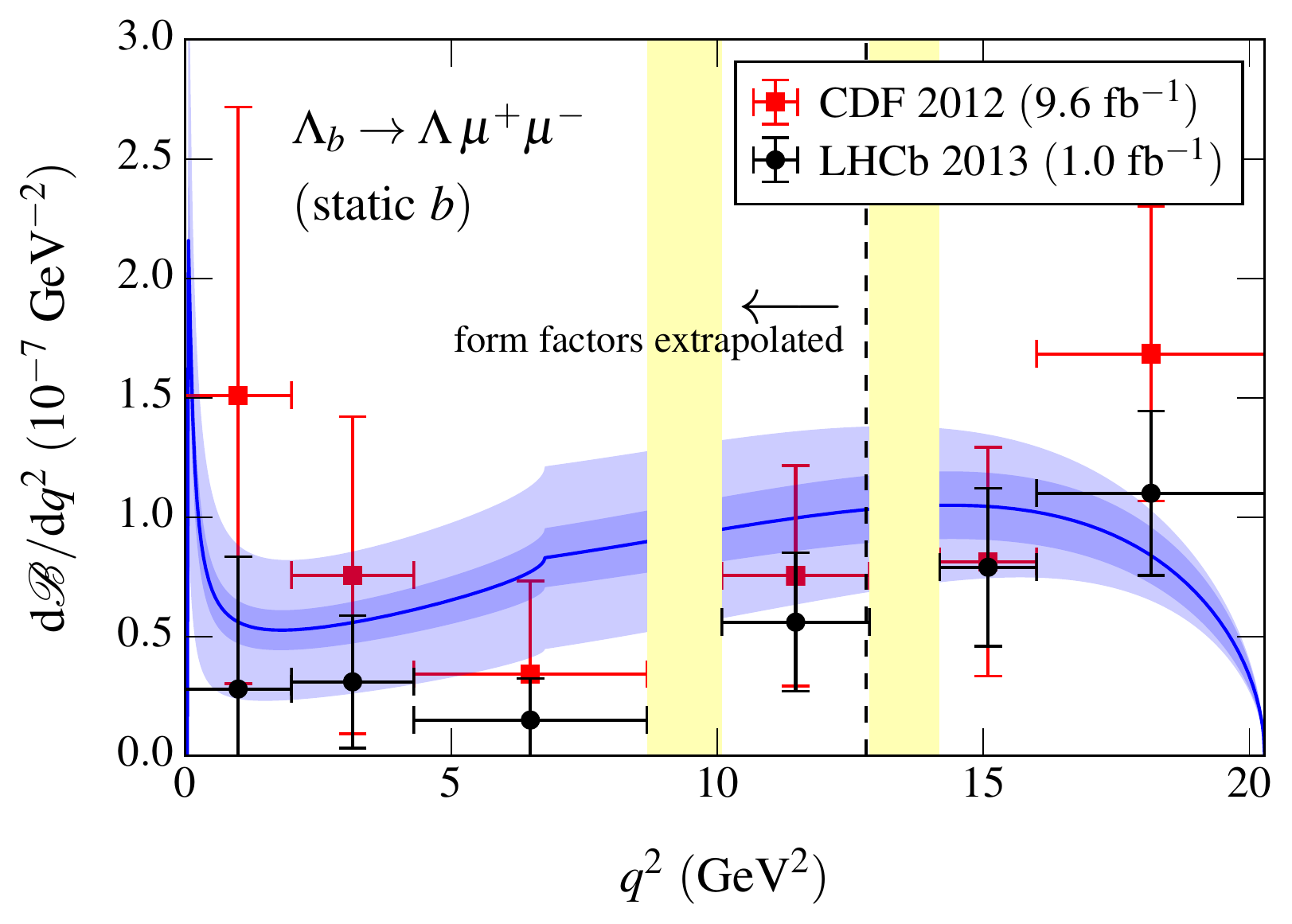} \vspace{-3ex}
\end{center}
\caption{\label{fig:LambdabLambdaBF}Differential branching fraction of $\Lambda_b \to \Lambda\, \mu^+ \mu^-$, calculated
in the static approximation for the $b$ quark. The inner error band originates from the uncertainties
in $F_\pm$, as well as the uncertainties in the non-lattice input parameters. The outer error band additionally includes the uncertainty
associated with the static approximation, which is of order $\sqrt{\Lambda_{\rm QCD}^2+|\mathbf{p}^\prime|^2}/m_b$. The vertical
yellow bands indicated the excluded regions around the $J/\psi$ and $\psi(2S)$ resonances, where the neglected long-distance
effects are dominant. The experimental results are from Refs.~\cite{CDF2012, Aaij:2013hna}.}
\end{figure}

\begin{figure}[!ht]
\begin{center}
\includegraphics[width=0.49\linewidth]{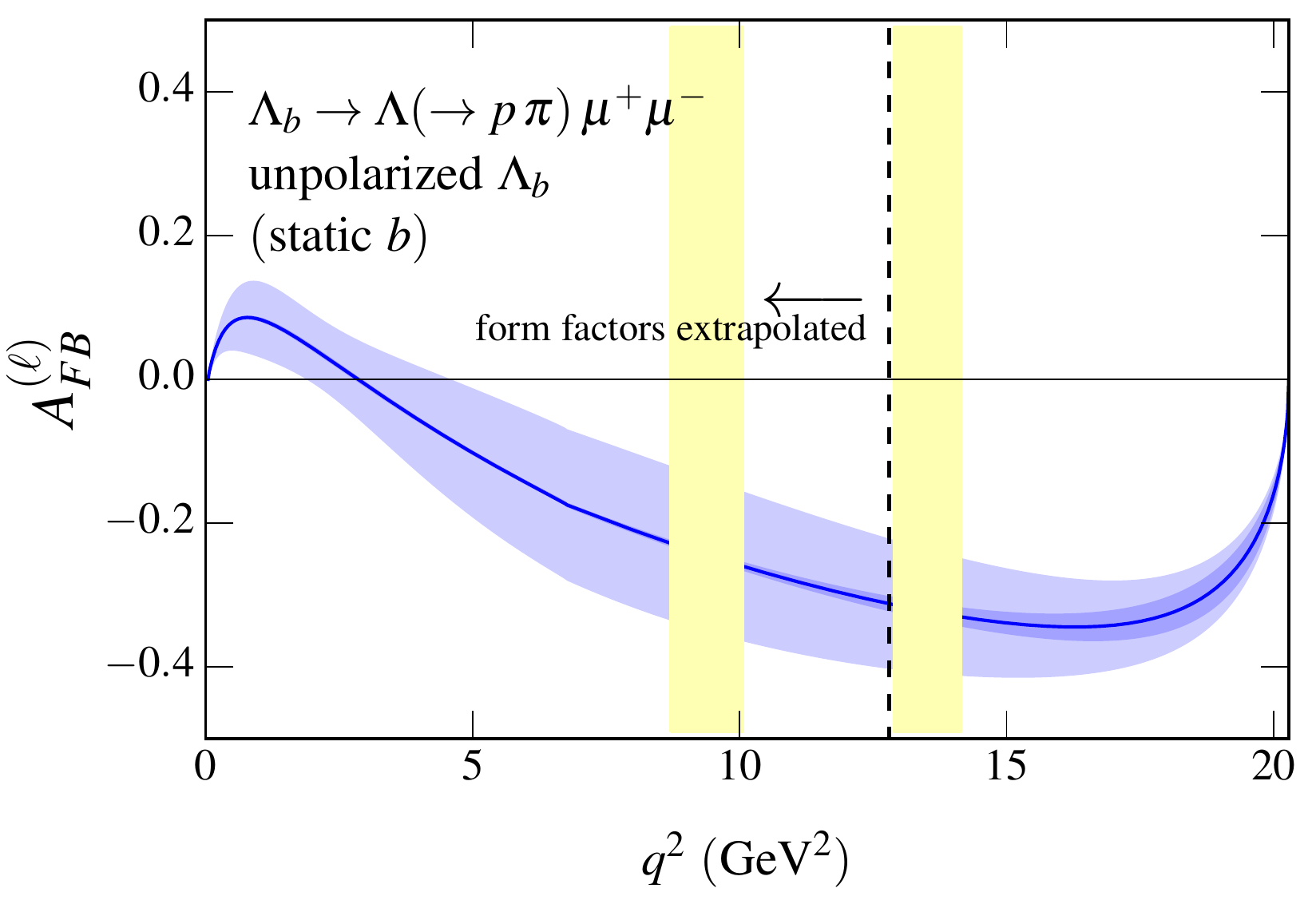} \hfill
\includegraphics[width=0.49\linewidth]{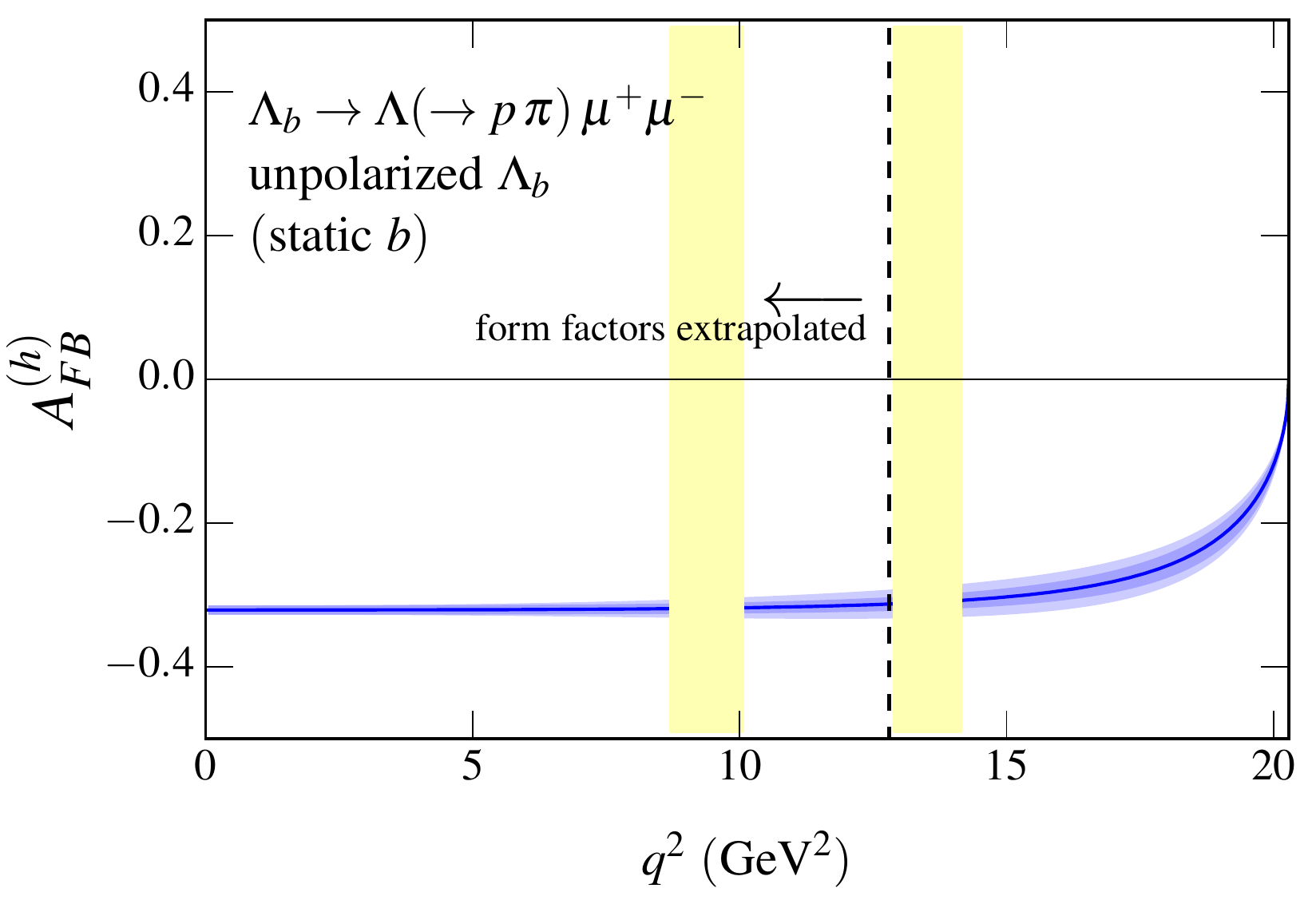} \vspace{-3ex}
\end{center}
\caption{\label{fig:LambdabLambdaAFB}Lepton-side (left) and hadron-side (right) forward-backward asymmetries of
$\Lambda_b \to \Lambda(\to p\,\pi) \, \mu^+ \mu^-$ with unpolarized $\Lambda_b$, calculated in the static approximation for the $b$ quark.}
\end{figure}

\begin{figure}[!ht]
\begin{center}
\includegraphics[width=0.55\linewidth]{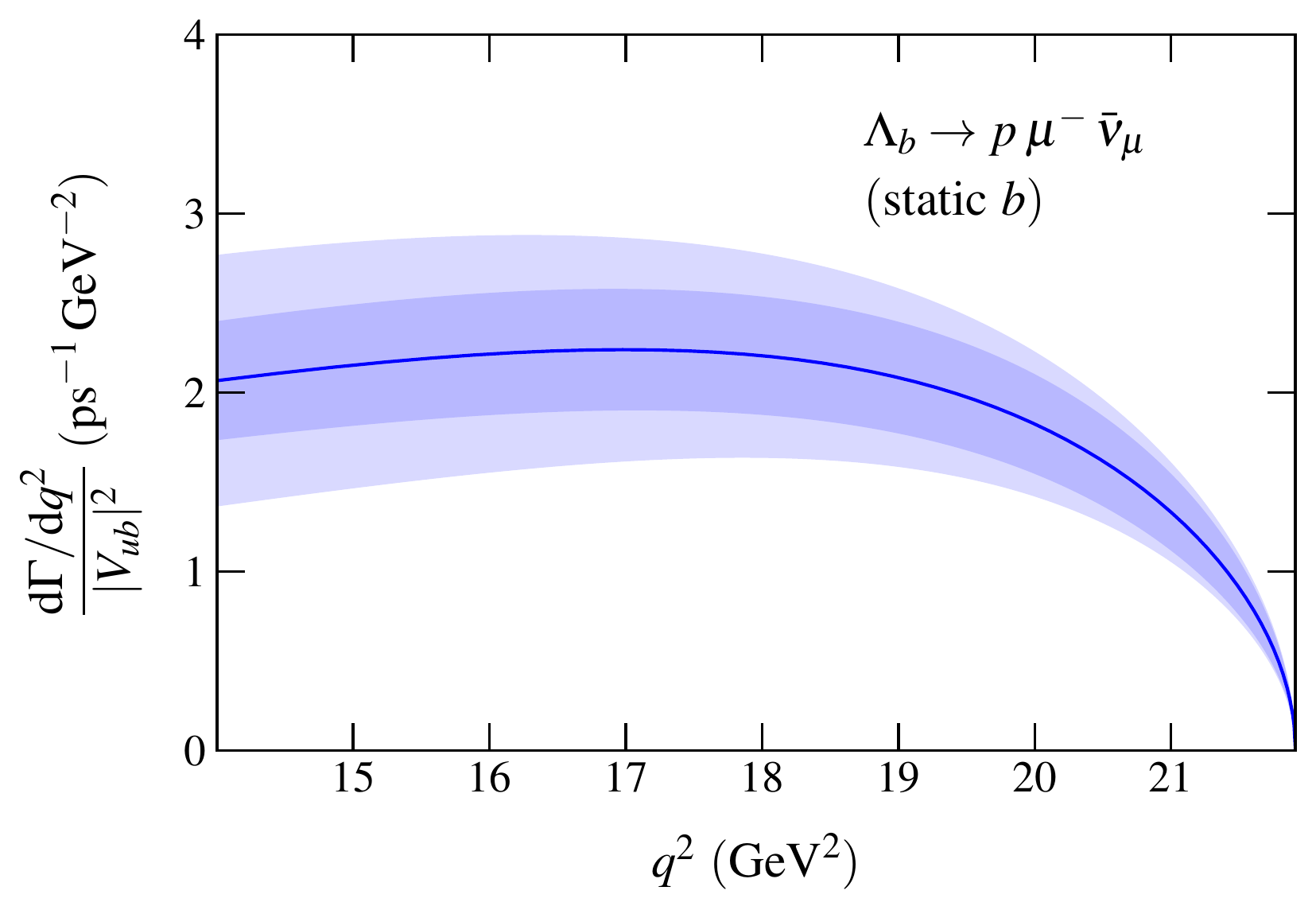} \vspace{-3ex}
\end{center}
\caption{\label{fig:Lambdabproton}$1/|V_{ub}|^2$ times the $\Lambda_b \to p \mu^- \bar{\nu}_\mu$ differential decay rate,
calculated in the static approximation for the $b$ quark \cite{Detmold:2013nia}.}
\end{figure}

In Fig.~\ref{fig:LambdabLambdaBF}, I show the $\Lambda_b \to \Lambda\,\mu^+\mu^-$ differential branching fraction
($\mathrm{d}\mathcal{B}/\mathrm{d}q^2=\tau_{\Lambda_b}\mathrm{d}\Gamma/\mathrm{d}q^2$), calculated in the Standard Model
with the form factors $F_\pm$ from Fig~\ref{fig:staticFFs}. Only the matrix elements of $O_7$, $O_9$, and $O_{10}$, which
can be expressed in terms of the form factors, are included, with some perturbative corrections to partially account for
the four-quark operators. These corrections fail to describe the strong nonperturbative enhancement of the matrix elements of $O_1$ and $O_2$
that is expected for $\sqrt{q^2}$ near the mass of a charmonium resonance with $J^{PC}=1^{--}$, so one has to stay away from these
regions. Besides the missing long-distance effects, the largest source of uncertainty in this calculation is the static
approximation (for $q^2 \lesssim 13\:\:{\rm GeV}^2$, there is an additional unquantified uncertainty associated with the
extrapolation of the form factors). Also shown in Fig.~\ref{fig:LambdabLambdaBF} are experimental results from CDF
\cite{CDF2012} and LHCb \cite{Aaij:2013hna}. These currently agree with our Standard Model calculation, but given the
large uncertainties, there is still room for possible new physics. The LHCb results are based only on the 2011 data,
and results with much higher statistics are forthcoming. With more data, an angular analysis of $\Lambda_b \to \Lambda (\to p\,\pi^-) \ell^+ \ell^-$
will also become possible. Figure \ref{fig:LambdabLambdaAFB} contains two examples of angular observables:
the lepton-side forward-backward asymmetry $A_{FB}^{(\ell)}$, and the hadron-side forward-backward asymmetry
$A_{FB}^{(h)}=a\: P_z^{(\Lambda)}$ [see Eq.~(\ref{eq:thetap})], both calculated for unpolarized $\Lambda_b$
(the polarization of the $\Lambda_b$'s produced at the LHC is expected to be weakly transverse; the currently available
measurement is consistent with zero \cite{Aaij:2013oxa}).

Finally, Fig.~\ref{fig:Lambdabproton} shows the predicted differential decay rate of the charged-current decay
$\Lambda_b \to p\,\mu^-\bar{\nu}_\mu$, in the kinematic region where the form factor shape is reliably determined
by the lattice data. Note that unlike $\Lambda_b \to \Lambda\,\mu^+\mu^-$, this decay is not affected by long-distance
effects. The integrated rate for $q^2> 14\:\:{\rm GeV}^2$ is equal to
\begin{equation}
 \frac{1}{|V_{ub}|^2}\int_{14\:{\rm GeV}^2}^{q^2_{\rm max}}
\frac{\mathrm{d}\Gamma (\Lambda_b \to p\: \mu^- \bar{\nu}_\mu)}{\mathrm{d}q^2} \mathrm{d} q^2
 \:=\: 15.3 \pm 2.4 \pm 3.4\:\: \rm{ps}^{-1}, \label{eq:Vubstatic}
\end{equation}
where the first uncertainty stems from the uncertainty in the form factors $F_+$ and $F_-$, and the second uncertainty
is due to the static approximation.

\section{\label{sec:relativistic}Lattice calculation with relativistic $b$ quarks}

The uncertainties in the calculations of $\Lambda_b$ decay observables in the previous section are dominated by the
$\mathcal{O}(\Lambda/m_b)$ errors caused by the static approximation. To eliminate these errors, I am currently performing
new calculations of the full set of relativistic form factors defined in Eqs.~(\ref{eq:fV})-(\ref{eq:fTA}), with a
``relativistic'' lattice action for the $b$ quark. These new calculations are done on the same RBC/UKQCD gauge field ensembles,
and reuse the existing domain-wall propagators for the $u$, $d$, and $s$ quarks (see Table \ref{tab:params}).
In addition to the $\Lambda_b \to \Lambda$ and $\Lambda_b \to p$ form factors, I now also compute the $\Lambda_b \to \Lambda_c$
form factors, which are relevant for the decay $\Lambda_b \to \Lambda_c \ell^- \bar{\nu}_\ell$. This decay can be used
to determine $|V_{cb}|$, and is a major background for the measurement of $\Lambda_b \to p \ell^- \bar{\nu}_\ell$ at LHCb.

The lattice actions for the $b$ and $c$ quarks are based on the Fermilab approach \cite{ElKhadra:1996mp}
and have the form
\begin{equation}
S_{\rm RHQ} = a^4 \sum_x \bar{Q} \left( m + \gamma_0 \nabla_0 - \frac{a}{2} \nabla_0^2 + {\xi} \bss{\gamma} \cdot \bss{\nabla}  - \frac{a}{2} {\xi} \bss{\nabla}^2
+ \frac{a}{2} {c_E}\: i\sigma^{0j} G_{0j} + \frac{a}{4} {c_B}\: i\sigma^{jk} G_{jk} \right) Q.
\end{equation}
The values of the parameters used here are given in Table \ref{tab:HQparams}. For the bottom quarks, I adopted the choice made by
RBC and UKQCD collaborations, who have tuned the three parameters $m$, $\xi$, and $c_E=c_B$ nonperturbatively \cite{Christ:2006us, Aoki:2012xaa}.
For the charm quarks, the two parameters $m$ and $\xi$ have been tuned nonperturbatively by Z.~S.~Brown, with $c_E$ and $c_B$
set to their tadpole-improved tree-level values \cite{Brown}.

\begin{table}
\begin{center}
\small
\begin{tabular}{ccccccccccc}
\hline\hline
Sets          & $a m^{(b)}$ & $\xi^{(b)}$ & $c_{E,\,B}^{(b)}$ & $Z_V^{(bb)}$   & $a m^{(c)}$ & $\xi^{(c)}$  &  $c_{E}^{(c)}$ & $c_{B}^{(c)}$  &  $Z_V^{(cc)}$  \\
\hline
$\mathtt{C*}$ & $8.45$      & $3.1\nb$    & $5.8\nb$          & $10.037(34)$   & $\wm0.1214$ & $1.2362$     &  $1.6650$      & $1.8409$       & $1.35695(38)$  \\
$\mathtt{F*}$ & $3.99$      & $1.93$      & $3.57$            & $5.270(13)\nb$ & $-0.0045$   & $1.1281$     &  $1.5311$      & $1.6232$       & $1.18343(26)$  \\
\hline\hline
\end{tabular}\vspace{-2ex}
\end{center}
\caption{\label{tab:HQparams} Parameters for the $b$ quark \cite{Aoki:2012xaa,Kawanai:2013qxa} and the $c$ quark \cite{Brown} on the coarse and fine lattices. }
\end{table}

The matching of the $b \to q$ currents ($q=u,s,c$) to the $\overline{\mathrm{MS}}$ scheme is now performed using the
``mostly nonperturbative'' method \cite{ElKhadra:2001rv}, writing
\begin{equation}
 J_\Gamma = \rho_\Gamma \,\sqrt{Z_V^{(qq)} Z_V^{(bb)}}  \Big[\, \bar{q}\,\Gamma b + a \sum_{i} c_{\Gamma,i}\:\, J_{\Gamma,i}^{(\nabla)} \Big], \label{eq:JGamma}
\end{equation}
where $Z_V^{(qq)}$ and $Z_V^{(bb)}$ are the renormalization factors of the currents $\bar{q}\gamma_0 q$ and $\bar{b}\gamma_0 b$, which
have been computed nonperturbatively \cite{Aoki:2010dy,Kawanai:2013qxa}. The remaining factors $\rho_\Gamma$ (which are expected to be close to unity)
and the $\mathcal{O}(a)$-improvement coefficients $c_{\Gamma,i}$ are being computed in tadpole-improved one-loop perturbation
theory using the framework PhySyHCAl \cite{Lehner:2012bt}. In the preliminary results presented in the following, the matching
coefficients $\rho_\Gamma$  are still missing for the tensor/pseudotensor currents and for all $b\to c$ currents.
Furthermore, only the tree level $\mathcal{O}(a)$-improvement is included.

\begin{figure}
\begin{center}
 \includegraphics[width=0.5\linewidth]{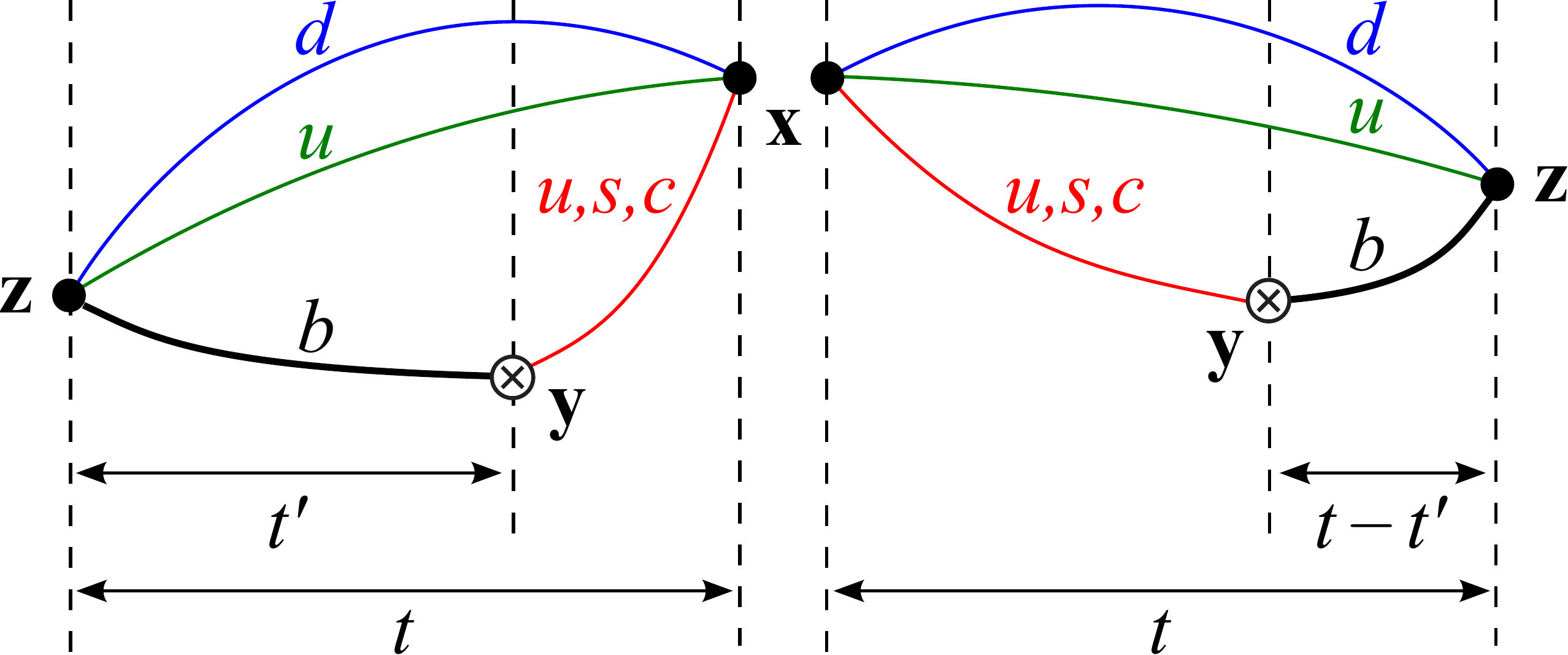}  \vspace{-3ex}
\end{center}
 \caption{\label{fig:threept}Forward (left) and backward (right) three-point functions.}
\end{figure}

To determine the form factors, I work in the $\Lambda_b$ rest frame and compute the ``forward'' and ``backward'' three-point
functions
\begin{eqnarray}
 C^{(3,\mathrm{fw})}_{\delta\alpha}(\Gamma,\:\mathbf{p'}, t, t') &=& \sum_{\mathbf{y},\mathbf{z}} e^{-i\mathbf{p'}\cdot(\mathbf{x}-\mathbf{y})}
 \Big\langle X_{\delta}(x_0,\mathbf{x})\:\:\:\: J_\Gamma^\dag(x_0-t+t',\mathbf{y})
 \:\:\:\: \bar{\Lambda}_{b\alpha} (x_0-t,\mathbf{z}) \Big\rangle, \hspace{4ex} \label{eq:threept} \\
C^{(3,\mathrm{bw})}_{\alpha\delta}(\Gamma,\:\mathbf{p'}, t, t-t') &=& \sum_{\mathbf{y},\mathbf{z}}
e^{-i\mathbf{p'}\cdot(\mathbf{y}-\mathbf{x})} \Big\langle \Lambda_{b\alpha}(x_0+t,\mathbf{z})\:\:\:\: J_\Gamma(x_0+t',\mathbf{y})
\:\:\:\: \bar{X}_{\delta} (x_0,\mathbf{x}) \Big\rangle, \label{eq:threeptbw}
\end{eqnarray}
where $X_\delta$ is the interpolating field of the $p$, $\Lambda$, or $\Lambda_c$, and $\mathbf{p}^\prime$ is its momentum.
As illustrated in Fig.~\ref{fig:threept}, Eqs.~(\ref{eq:threept}) and (\ref{eq:threeptbw}) can be constructed from standard
shell-source light-, strange-, and charm-quark propagators with source location $(x_0,\mathbf{x})$; only the $b$-quark propagators
are sequential and need to be recomputed for each source-sink separation, $t$.

\begin{figure}
\begin{center}
 \includegraphics[width=\linewidth]{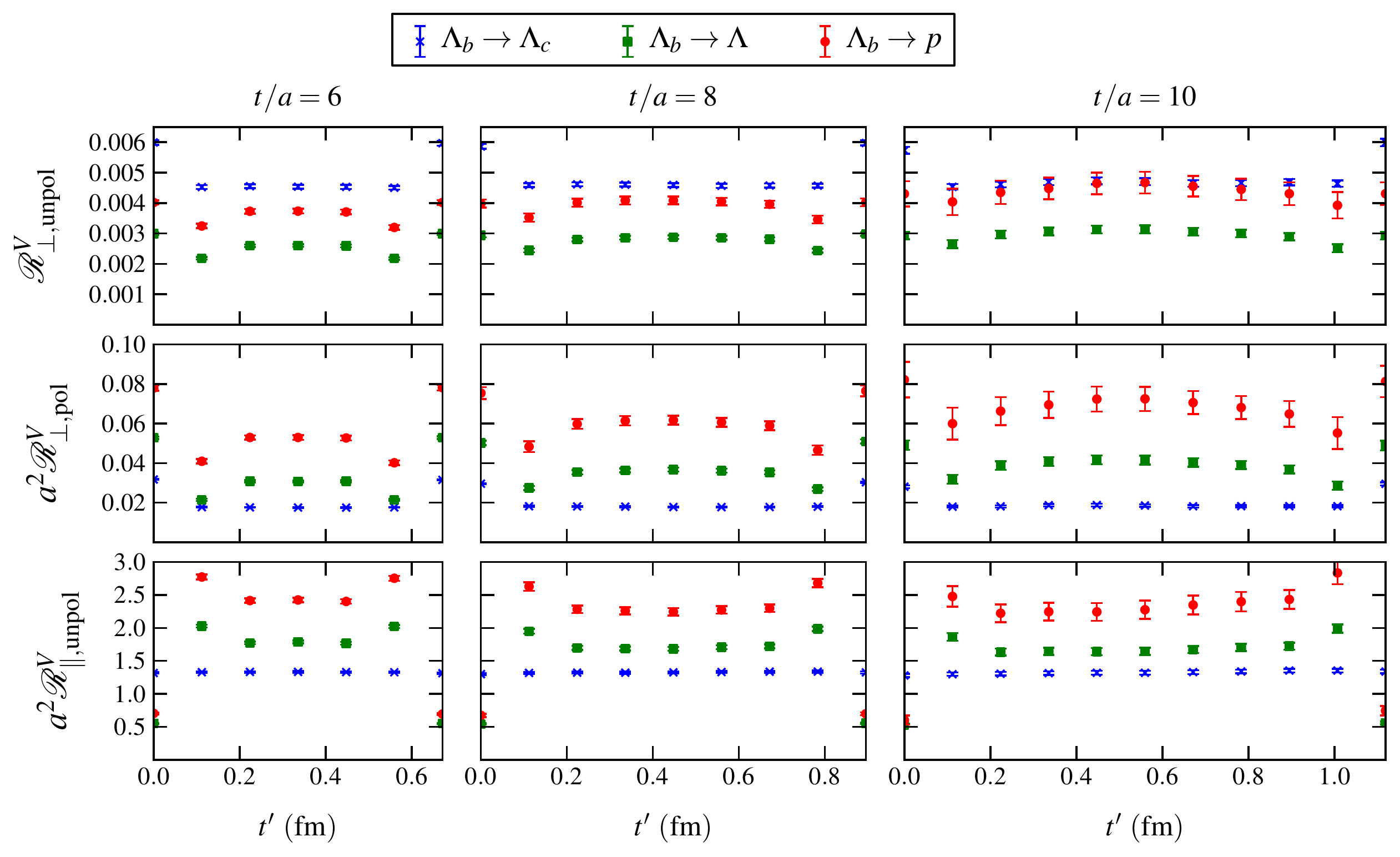}\vspace{-2ex}
 \caption{\label{fig:ratios}Preliminary results for the ratios (\protect\ref{eq:RperpUnpol}), (\protect\ref{eq:RperpPol}), and (\protect\ref{eq:RparUnpol})
 at $|\mathbf{p}^\prime|^2=3(2\pi/L)^2$, plotted for three different source-sink separations $t$. The data shown here are from the $\mathtt{C24}$ data set.}
\end{center}
\end{figure}

\begin{figure}
\begin{center}
 \includegraphics[width=0.37\linewidth]{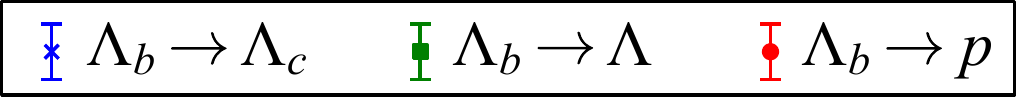}
 \vspace{1ex}

 \includegraphics[width=0.49\linewidth]{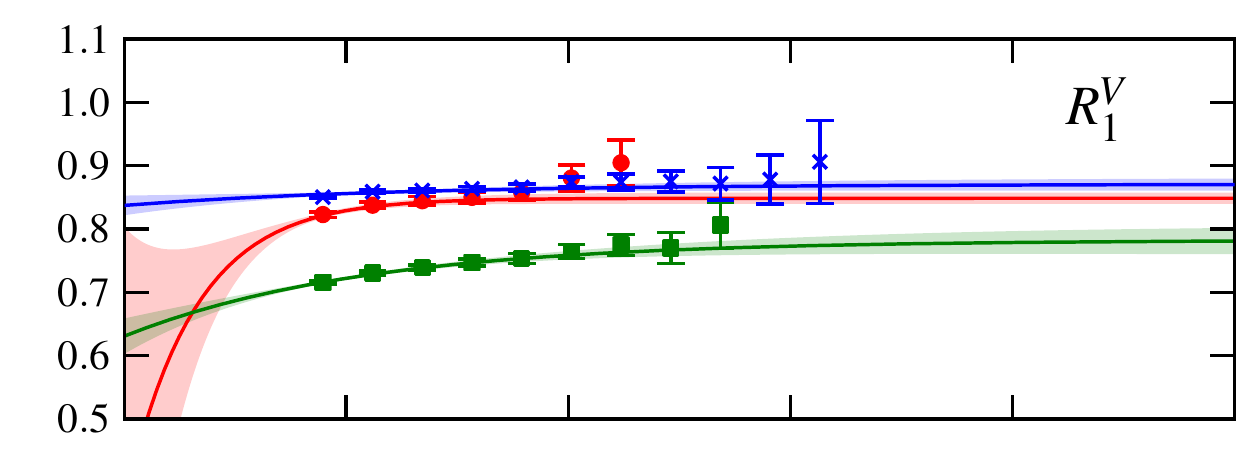} \hfill
 \includegraphics[width=0.49\linewidth]{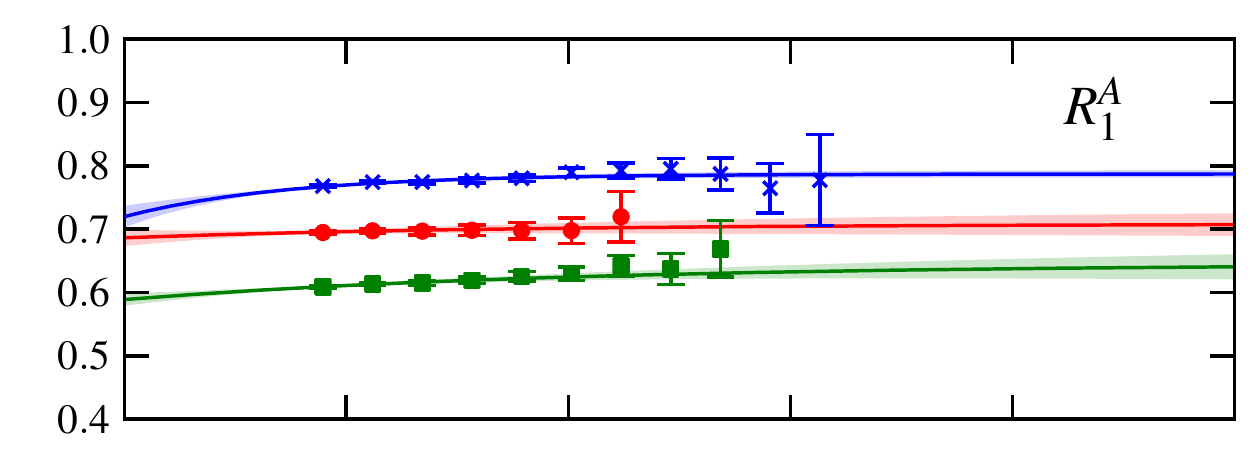} \\

 \includegraphics[width=0.49\linewidth]{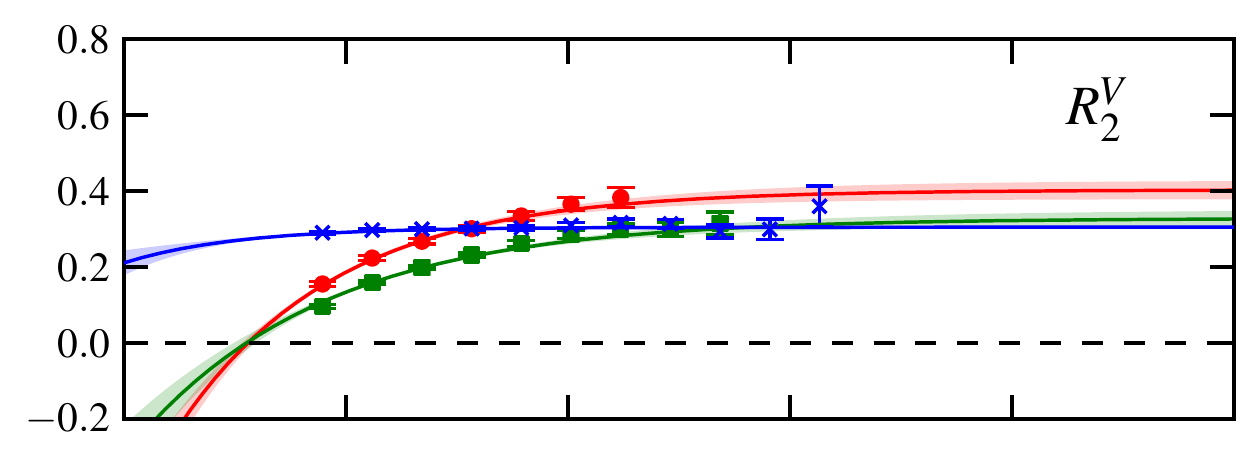} \hfill
 \includegraphics[width=0.49\linewidth]{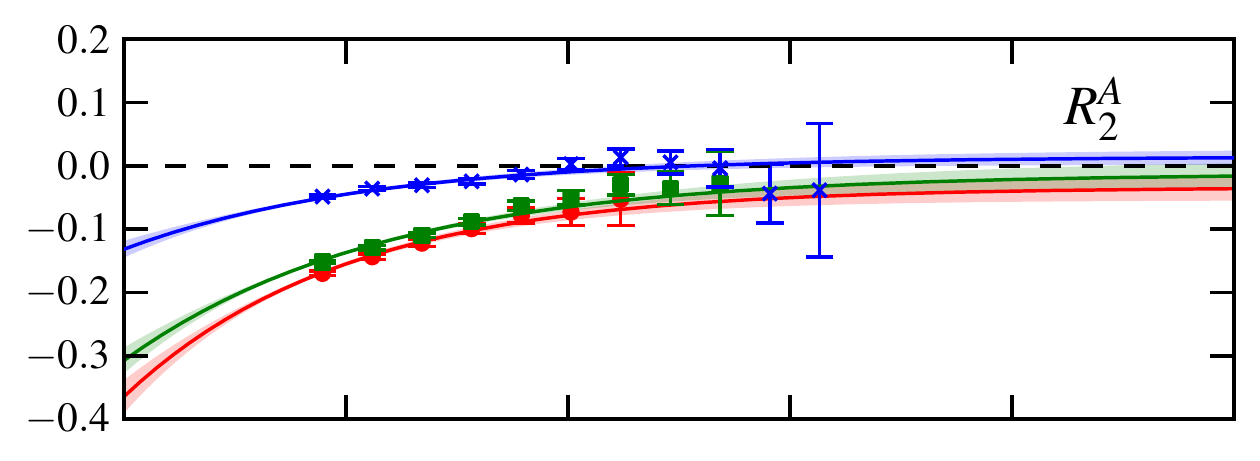} \\

 \includegraphics[width=0.49\linewidth]{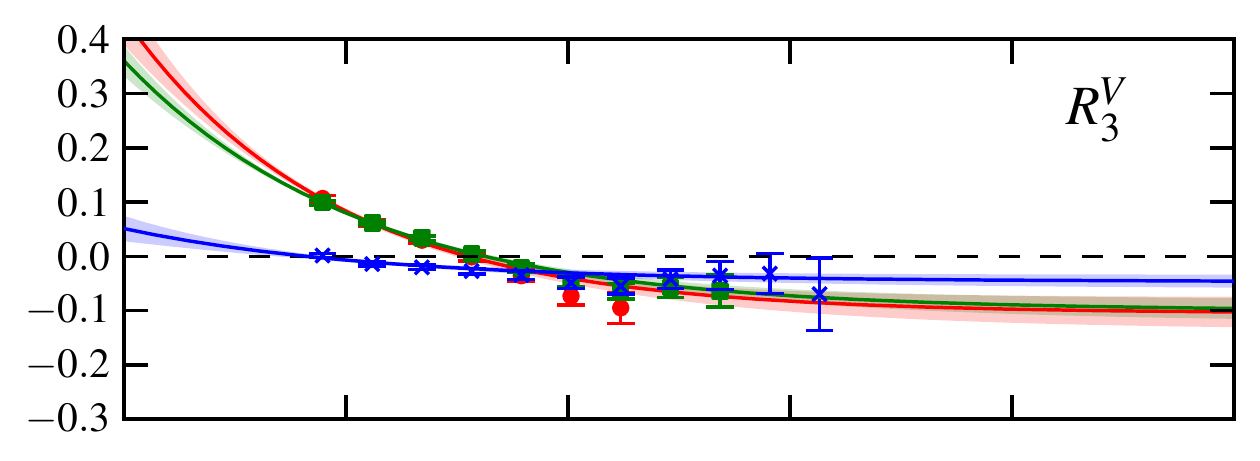} \hfill
 \includegraphics[width=0.49\linewidth]{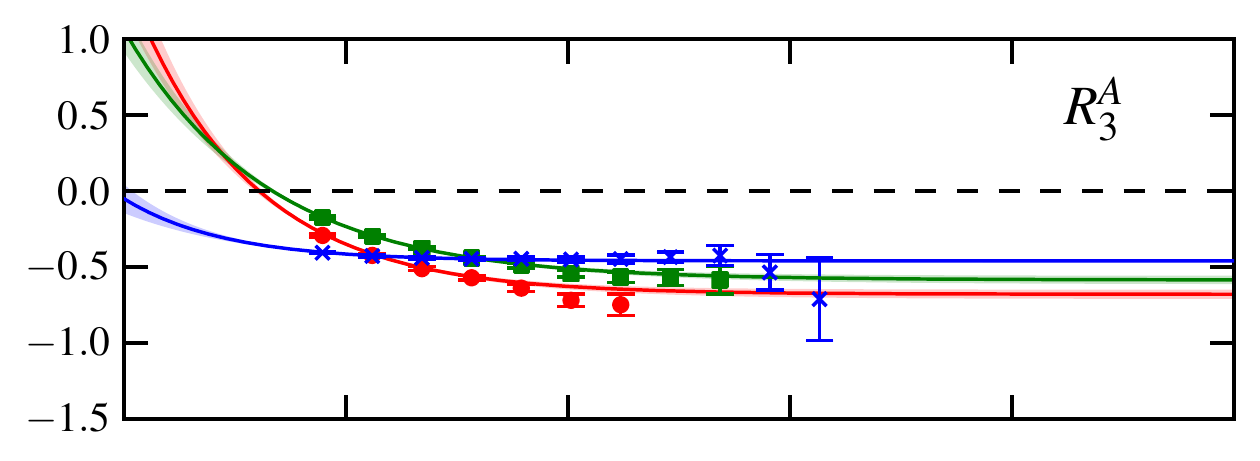} \\

 \includegraphics[width=0.49\linewidth]{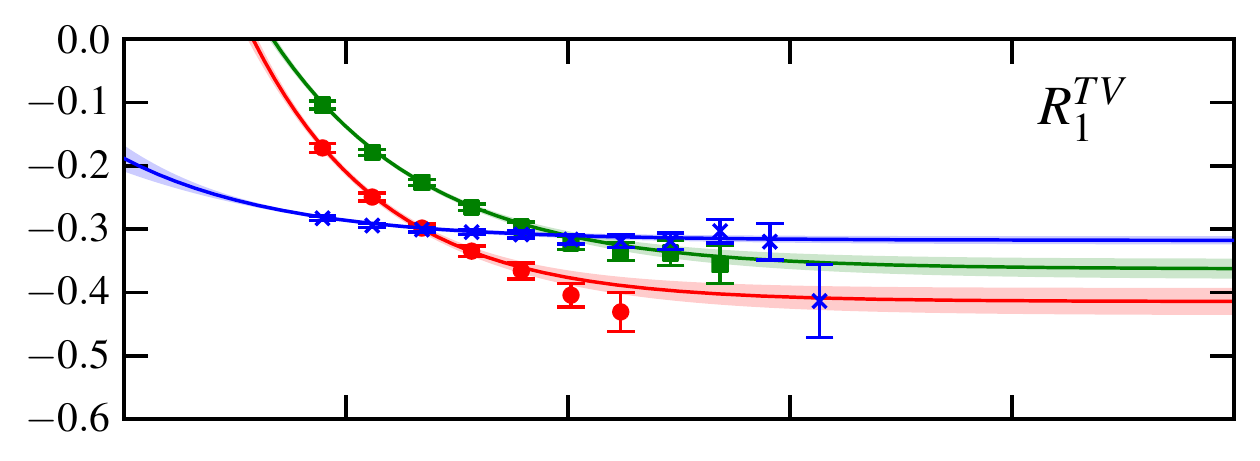} \hfill
 \includegraphics[width=0.49\linewidth]{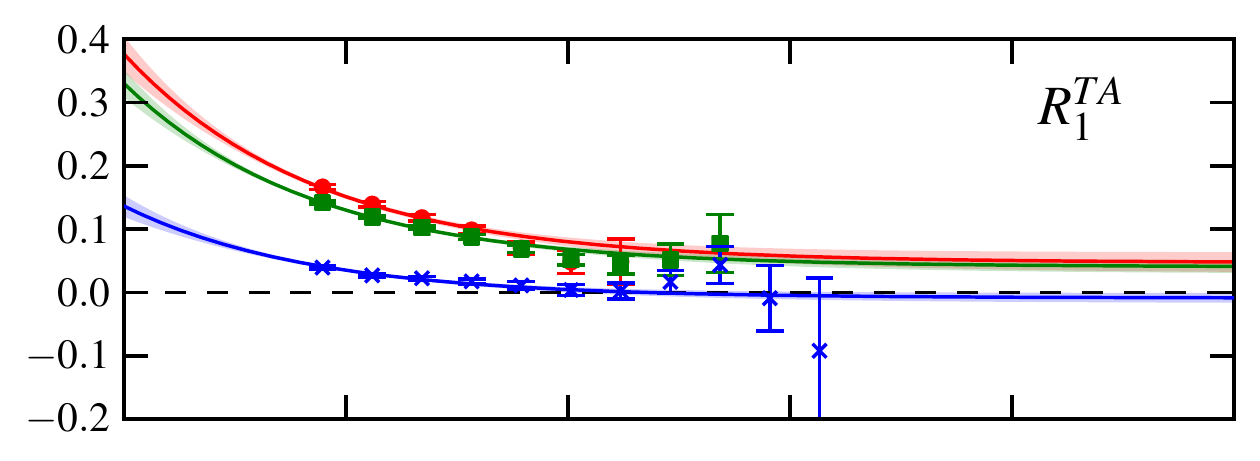} \\

 \includegraphics[width=0.49\linewidth]{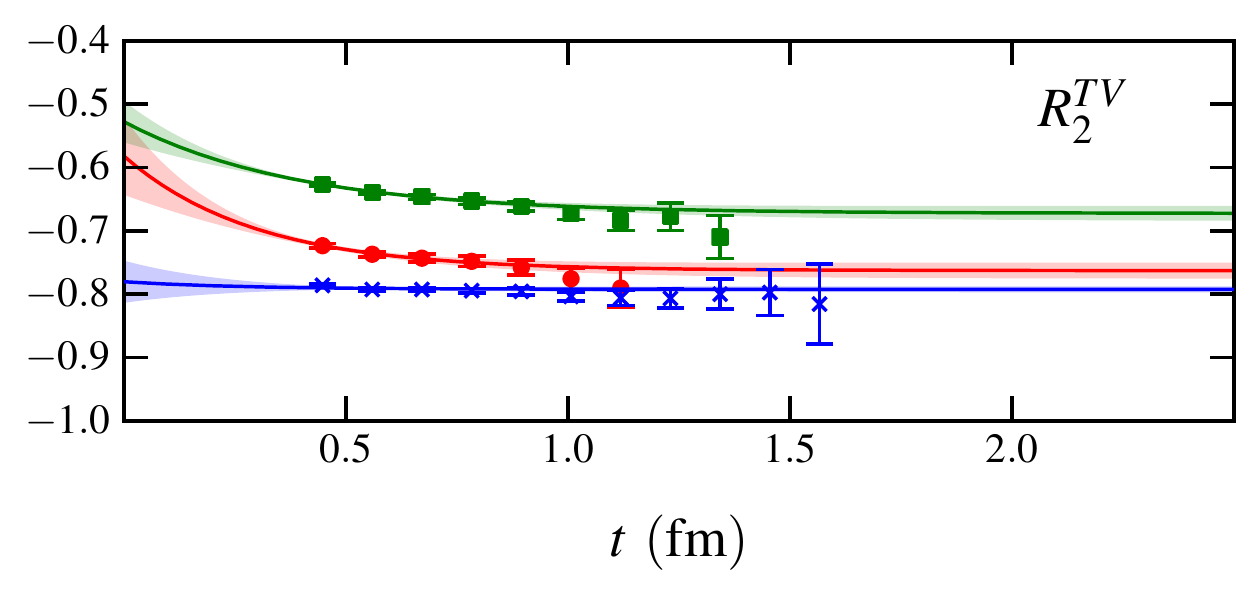} \hfill
 \includegraphics[width=0.49\linewidth]{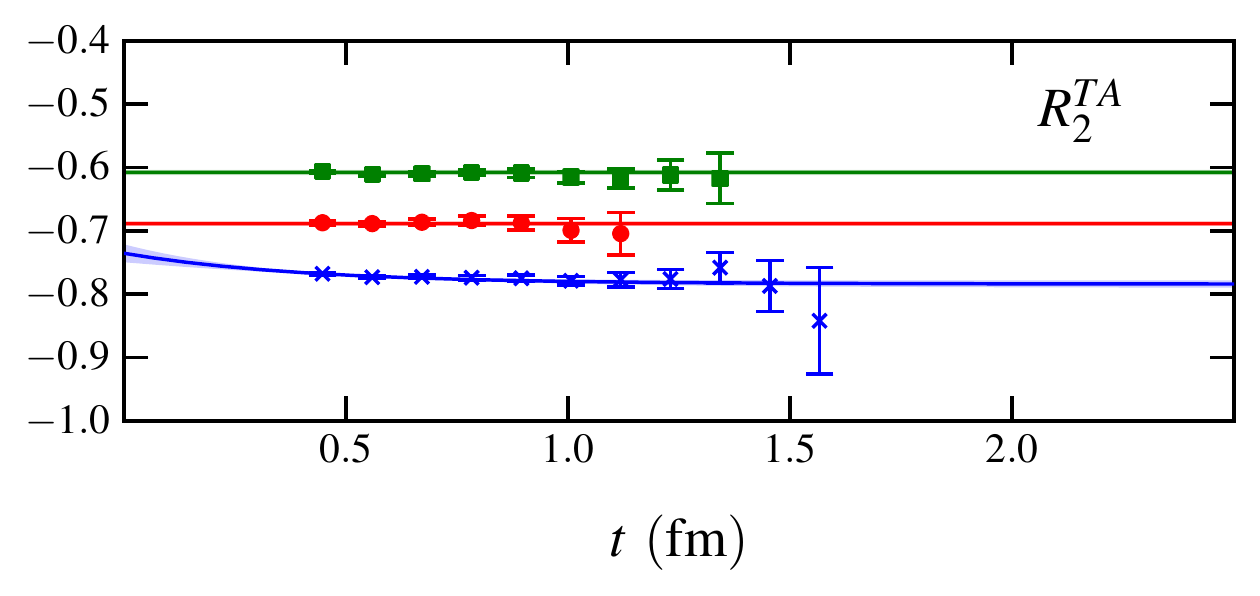} \\
 
 \vspace{-1ex}

 \caption{\label{fig:R}Preliminary results for the quantities $R_{1,2,3}^V$ [defined in Eqs.~(\protect\ref{eq:R1V})-(\protect\ref{eq:R3V})]
 and the analogous quantities $R_{1,2,3}^A$, $R_{1,2}^{TV}$, and $R_{1,2}^{TA}$, along with fits of the $t$-dependence
 ($t$ is the source-sink separation). The results shown here are at $|\mathbf{p}^\prime|^2=3(2\pi/L)^2$ and are from the $\mathtt{C24}$ data set. }
\end{center}
\end{figure}

As an example, I will explain in the following how I extract the vector form factors $f_1^V$, $f_2^V$, and $f_3^V$ from the
correlation functions. The expressions for the other form factors are very similar. I first compute the following ratios
of three-point and two-point functions,
\begin{eqnarray}
\mathcal{R}_{\perp,\rm unpol}^V(\mathbf{p'},t,t') &=& \frac{ r_\mu[(1,\mathbf{0})] \: r_\nu[(1,\mathbf{0})] \:
\mathrm{Tr}\big[   C^{(3,{\rm fw})}(\mathbf{p'},\:\gamma^\mu, t, t') \:    C^{(3,{\rm bw})}(\mathbf{p'},\:\gamma^\nu, t, t-t')  \big] }
{\mathrm{Tr}\big[C^{(2,X)}(\mathbf{p'}, t)\big] \mathrm{Tr}\big[C^{(2,\Lambda_b)}(t)\big] }, \label{eq:RperpUnpol} \\
\nonumber \mathcal{R}_{\perp,\rm pol}^V(\mathbf{p'},t,t') &=& r_\mu[(0,\mathbf{e}_j\times \mathbf{p}')] \:   r_\nu[(0,\mathbf{e}_k\times \mathbf{p}')] \\
&& \times \frac{  \:
\mathrm{Tr}\big[  C^{(3,{\rm fw})}(\mathbf{p'},\:\gamma^\mu, t, t') \gamma_5 \gamma^j \:    C^{(3,{\rm bw})}(\mathbf{p'},\:\gamma^\nu, t, t-t') \gamma_5 \gamma^k  \big] }
{\mathrm{Tr}\big[C^{(2,X)}(\mathbf{p'}, t)\big] \mathrm{Tr}\big[C^{(2,\Lambda_b)}(t)\big] },  \label{eq:RperpPol}  \\
\mathcal{R}_{\parallel,\rm unpol}^V(\mathbf{p'},t,t') &=& \frac{ q_\mu \: q_\nu \:
\mathrm{Tr}\big[   C^{(3,{\rm fw})}(\mathbf{p'},\:\gamma^\mu, t, t') \:    C^{(3,{\rm bw})}(\mathbf{p'},\:\gamma^\nu, t, t-t')  \big] }
{\mathrm{Tr}\big[C^{(2,X)}(\mathbf{p'}, t)\big] \mathrm{Tr}\big[C^{(2,\Lambda_b)}(t)\big] },  \label{eq:RparUnpol}
\end{eqnarray}
where $q=p-p'$ and $r[n]=n-\frac{(q\cdot n)}{q^2}q$. In these ratios, all overlap factors as well as the $t$ and $t^\prime$ dependence cancel for the
ground-state contribution. Furthermore, these ratios are rotationally symmetric (in the continuum), and I average them over the direction of
$\mathbf{p}^\prime$. I computed the ratios for all source-sink separations in the range $4 \leq t/a \leq 15$ on the coarse lattices and $5 \leq t/a \leq 15$ on the fine lattices,
which gives excellent control over excited-state contamination. Examples of numerical results for Eqs.~(\ref{eq:RperpUnpol}), (\ref{eq:RperpPol}),
and (\ref{eq:RparUnpol}) are shown in Fig.~\ref{fig:ratios}.
I then calculate the functions
\begin{eqnarray}
 R_1^V &=& \sqrt{\frac{E_X }{E_X+m_X}} \frac{q^2 }{(E_X-m_X) \left((m_{\Lambda_b}+m_X)^2-q^2\right)}
 \left[2 (m_{\Lambda_b}+m_X) \sqrt{ \mathcal{R}_{\perp,\rm unpol}^V}-\sqrt{ \mathcal{R}_{\perp,\rm pol}^V}\,\right], \label{eq:R1V} \\
\nonumber R_2^V &=& \sqrt{\frac{E_X }{E_X+m_X}} \frac{m_{\Lambda_b} }{(E_X-m_X) \left((m_{\Lambda_b}+m_X)^2-q^2\right)}
\left[(m_{\Lambda_b}+m_X) \sqrt{ \mathcal{R}_{\perp,\rm pol}^V}-2 q^2 \sqrt{\mathcal{R}_{\perp,\rm unpol}^V}\,\right], \label{eq:R2V} \\
&& \\
 R_3^V &=& \frac{2 m_{\Lambda_b} \sqrt{ \frac{E_X}{E_X+m_X} \mathcal{R}_{\parallel,\rm unpol}^V }
 -m_{\Lambda_b} (m_{\Lambda_b}-m_X)R_1^V }{ m_{\Lambda_b} (m_{\Lambda_b}-2 E_X)+m_X^2 },  \label{eq:R3V}
\end{eqnarray}
where I evaluate $\mathcal{R}_{\perp,\rm unpol}^V$, $\mathcal{R}_{\perp,\rm pol}^V$, and $\mathcal{R}_{\parallel,\rm unpol}^V$
at the midpoint $t'=t/2$ [in Eq.~(\ref{eq:R3V}), $R_1^V$ is given by Eq.~(\ref{eq:R1V})]. Up to excited-state contamination
that decays exponentially for $t\to\infty$, $R_1^V$, $R_2^V$, and $R_3^V$ are equal to the form factors $f_1^V$, $f_2^V$,
and $f_3^V$, respectively. Examples of numerical results for $R_1^V$, $R_2^V$, $R_3^V$ and the analogous quantities for
the other seven form factors are shown in Fig.~\ref{fig:R}. To extract the ground-state contributions, I fit the dependence
on the source-sink separation using
\begin{equation}
 R(t)=f + A \: e^{-\delta\:t},\hspace{2ex}\delta=\delta_{\rm min} + e^{\,l}\:\:{\rm GeV}
\end{equation}
with separate parameters $f$, $A$, and $l$ for each form factor, each value of $|\mathbf{p}^\prime|^2$, and each data set.
Here, $\delta_{\rm min}$ is a small minimum energy gap, introduced for numerical stability. The fits include constraints
that limit the variation of the parameters $l$ across the different data sets (for a given form factor and given value of
$|\mathbf{p}^\prime|^2$). These fits are also shown in Fig.~\ref{fig:R}. Some of the ground-state form factors are in fact
close to zero, in which case the careful removal of excited-state contamination is particularly important.

\begin{figure}
\begin{center}
 \framebox{\scriptsize \tt \bfseries \textcolor{red}{C14}\hspace{2ex} \textcolor{magenta}{C24}\hspace{2ex}
 \textcolor{orange}{C54}\hspace{2ex} \textcolor{brown}{C53}\hspace{2ex} \textcolor{darkgreen}{F23}\hspace{2ex}
 \textcolor{darkblue}{F43}\hspace{2ex} \textcolor{black}{F63}} \\
 \includegraphics[width=\linewidth]{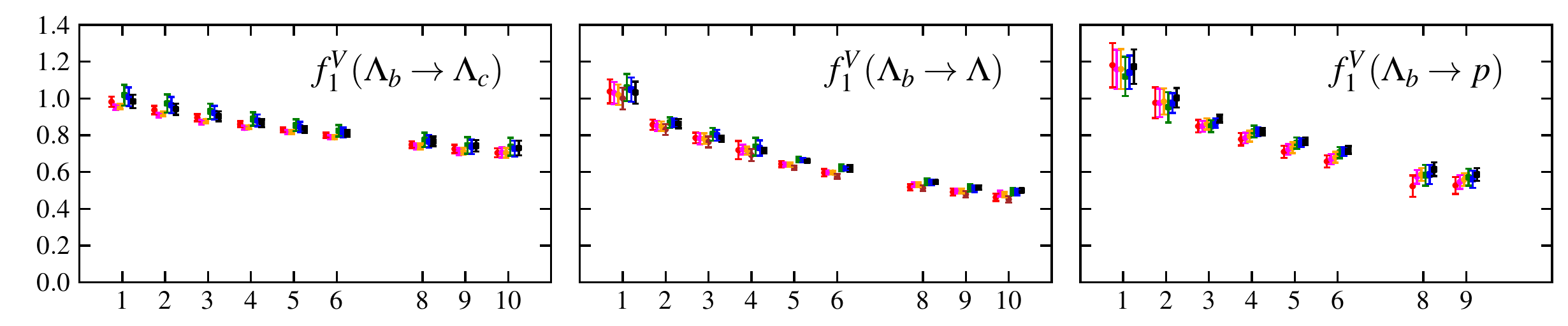} \\ \vspace{-0.5ex}
 \includegraphics[width=\linewidth]{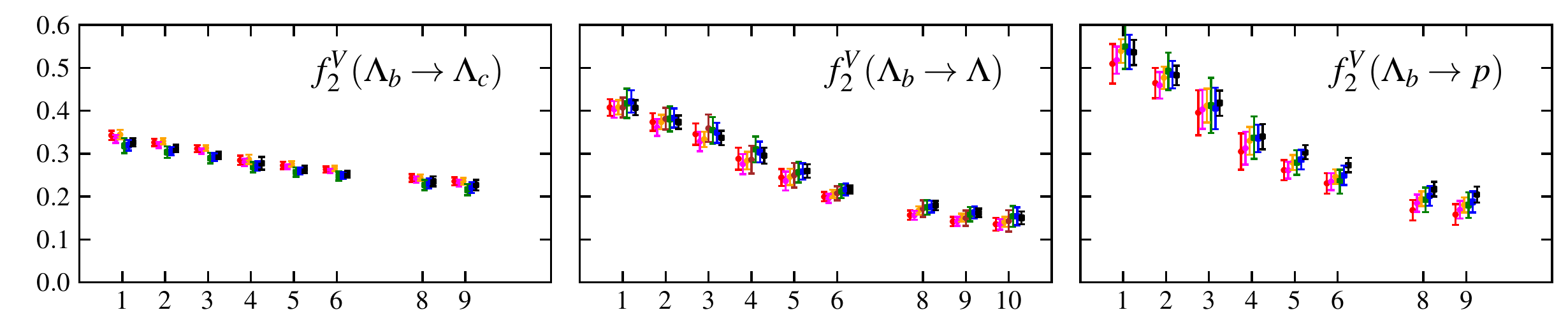} \\ \vspace{-0.5ex}
 \includegraphics[width=\linewidth]{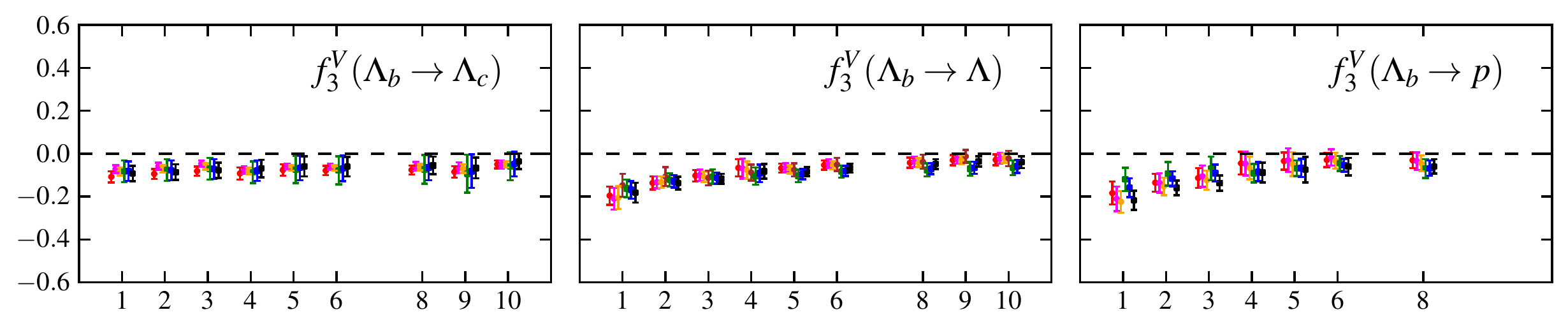} \\ \vspace{-0.5ex}
 \includegraphics[width=\linewidth]{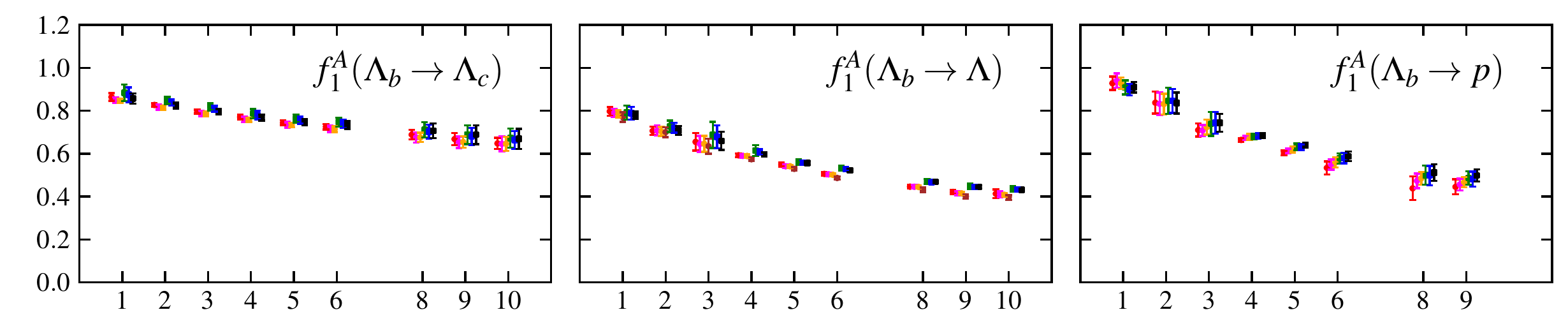} \\ \vspace{-0.5ex}
 \includegraphics[width=\linewidth]{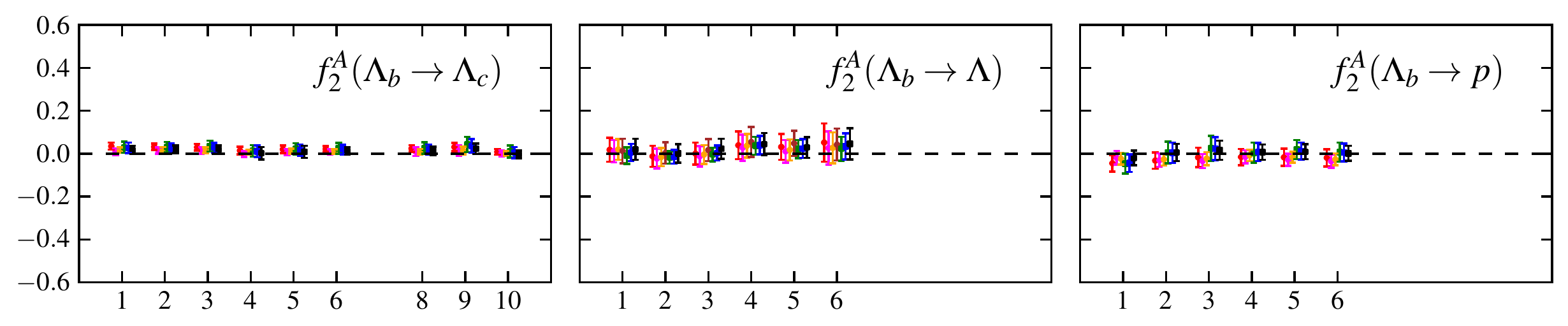} \\ \vspace{-0.5ex}
 \includegraphics[width=\linewidth]{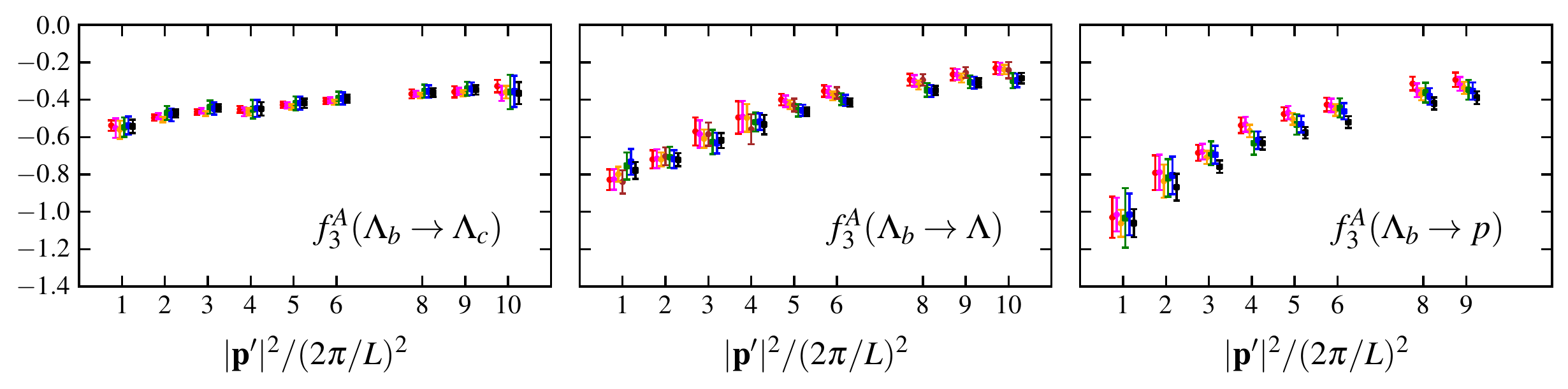} 
 \caption{\label{fig:vaff}Preliminary results for the form factors $f_1^V$, $f_2^V$, $f_3^V$, $f_1^A$, $f_2^A$, and $f_3^A$,
 plotted as a function of  $|\mathbf{p}^\prime|^2/(2\pi/L)^2$. The results from the different data sets are plotted with
 different colors and  are offset horizontally for clarity. The renormalization and $\mathcal{O}(a)$ improvement are still
 incomplete, as explained below Eq.~(\protect\ref{eq:JGamma}).}
\end{center}
\end{figure}

\begin{figure}
\begin{center}
 \framebox{\scriptsize \tt \bfseries \textcolor{red}{C14}\hspace{2ex} \textcolor{magenta}{C24}\hspace{2ex}
 \textcolor{orange}{C54}\hspace{2ex} \textcolor{brown}{C53}\hspace{2ex} \textcolor{darkgreen}{F23}\hspace{2ex}
 \textcolor{darkblue}{F43}\hspace{2ex} \textcolor{black}{F63}} \\
 \includegraphics[width=\linewidth]{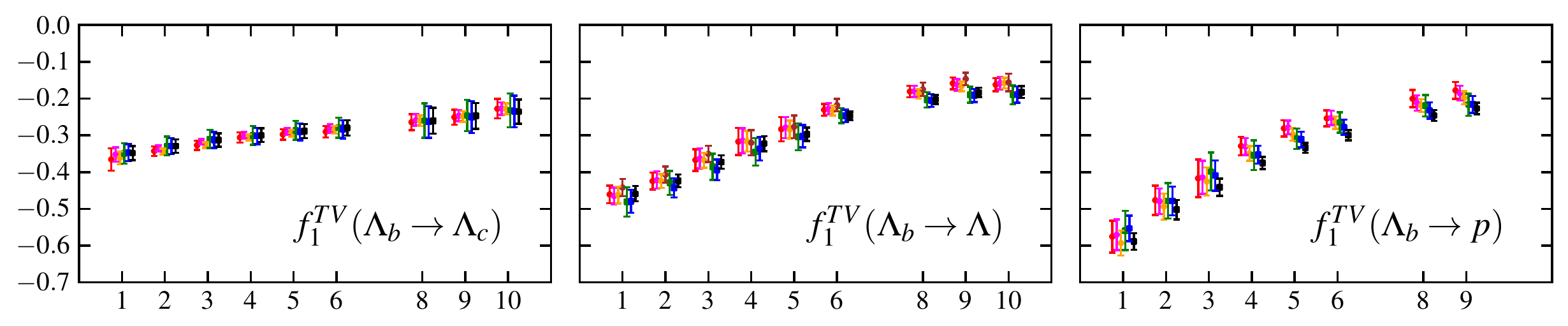} \\ \vspace{-0.5ex}
 \includegraphics[width=\linewidth]{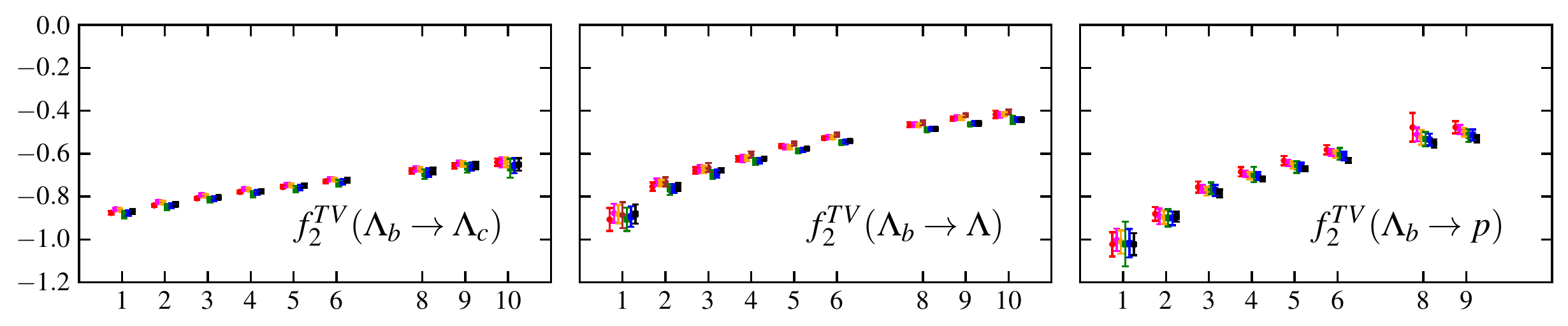} \\ \vspace{-0.5ex}
 \includegraphics[width=\linewidth]{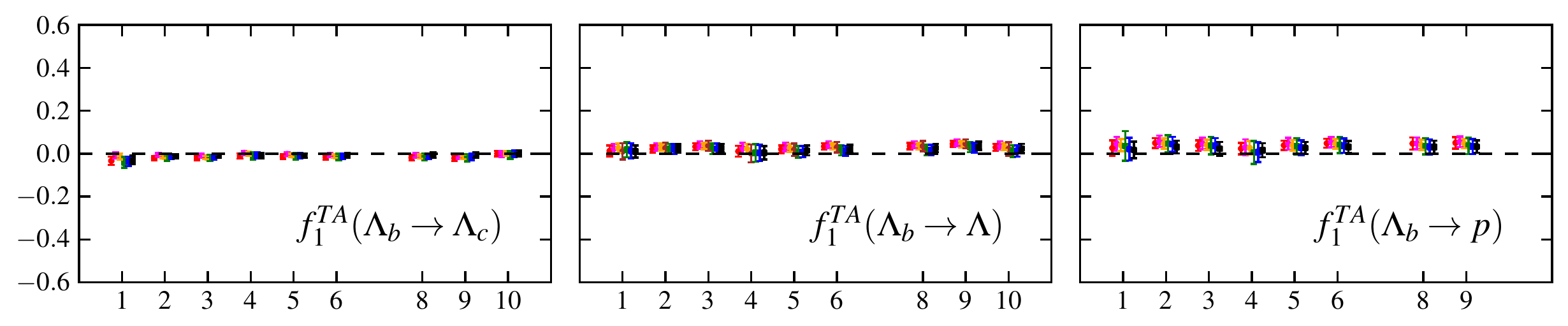} \\ \vspace{-0.5ex}
 \includegraphics[width=\linewidth]{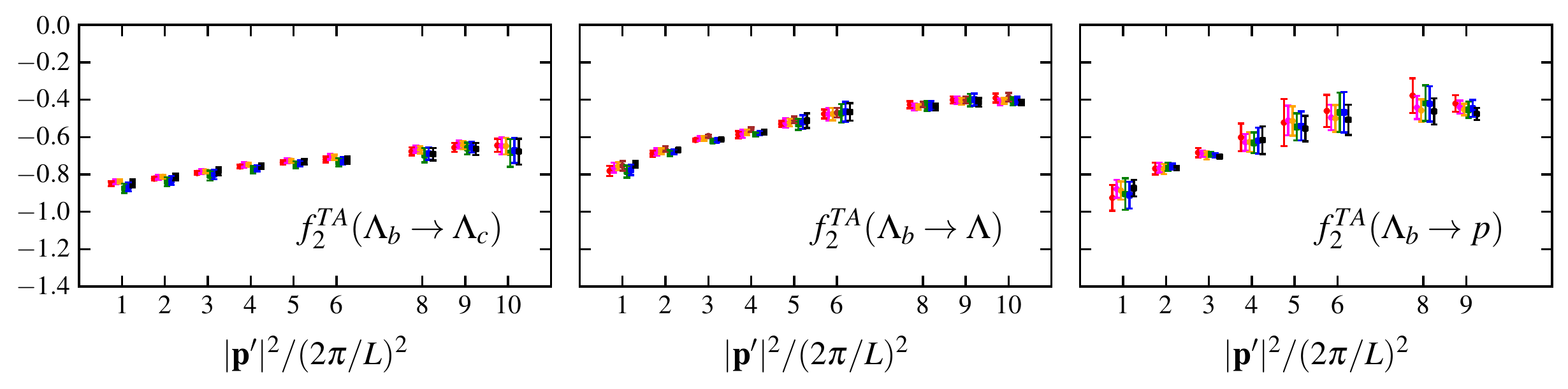}
 \caption{\label{fig:tvtaff}Like Fig.~\protect\ref{fig:vaff}, but for the form factors $f_1^{TV}$, $f_2^{TV}$, $f_1^{TA}$, and $f_2^{TA}$.}
\end{center}
\end{figure}

The preliminary results for all $\Lambda_b \to p$, $\Lambda_b \to \Lambda$, and $\Lambda_b \to \Lambda_c$ form factors
(still at nonzero lattice spacing and unphysical light-quark masses), are shown in Figs.~\ref{fig:vaff} and \ref{fig:tvtaff}.
The last step of the data analysis (after the renormalization and $\mathcal{O}(a)$-improvement are finalized)
will be to fit the form factor shapes and perform chiral and continuum extrapolations.

\section{\label{sec:outlook}Conclusions}

The $\Lambda_b$ baryons produced at the LHC in large quantities offer exciting new opportunities for
flavor physics, and lattice QCD calculations of $\Lambda_b$ decay form factors (and other quantities,
such distribution amplitudes \cite{Braun:2008ur}) are called for. The decay $\Lambda_b \to \Lambda \mu^+ \mu^-$
has excellent potential for constraining the Wilson coefficients $C_{7,9,10}$ and $C_{7,9,10}^\prime$,
which is especially interesting in light of recent hints of deviations from the Standard Model in
$B \to K^* \mu^+ \mu^-$ and $B_s \to \phi \mu^+ \mu^-$ \cite{BKstarAnomaly}. Furthermore, the analysis of the
decay $\Lambda_b \to p \mu^- \bar{\nu}_\mu$ will likely yield the first determination of the CKM matrix element
$|V_{ub}|$ at the LHC. Unlike the rare $b \to s$ decays, this process is not affected by long-distance contributions,
and the theory precision is limited only by the knowledge of the $\Lambda_b \to p$ form factors. Using the new calculations
of the relativistic form factors presented here, the uncertainty of the $\Lambda_b \to p \mu^- \bar{\nu}_\mu$
decay rate in the large-$q^2$ region will likely be reduced by at least a factor of 2 [compared to Eq.~(\ref{eq:Vubstatic})],
which will allow a $|V_{ub}|$ determination with a theory uncertainty of about 7\% or lower (the discrepancy between the
existing $|V_{ub}|$ extractions from inclusive and exclusive $B$-meson decays is about 30\% \cite{Beringer:1900zz}).

In the lattice calculations of the form factors, the careful removal of excited-state contamination is absolutely
essential, as can be seen in Fig.~\ref{fig:R}. This aspect is in fact under better control here than in typical
lattice calculations of nucleon form factors, because the three-point functions have been computed for a wide range
of source-sink separations. This is computationally affordable because no sequential light-quark propagators are needed
(cf.~Fig.~\ref{fig:threept}).

Now that the $\mathcal{O}(\Lambda/m_b)$ errors associated with the static approximation have been removed, other sources
of uncertainty are more important. These include the light-quark mass extrapolations of the form factors, for which
chiral perturbation theory is only of limited use \cite{Detmold:2013nia}. It is planned to perform new calculations directly
at the physical point, which will eliminate this uncertainty.

\vspace{1ex}

\noindent \textbf{Acknowledgments}: I thank William Detmold, Christoph Lehner, C.-J.~David Lin, and Matthew Wingate for useful discussions.
This work is supported by the U.S.~Department of Energy under cooperative research agreement Contract Number DE-FG02-94ER40818.
Numerical computations were performed using resources at NERSC and XSEDE resources at NICS and TACC.

\linespread{0.9}

\end{document}